\documentclass{article}
\usepackage[preprint]{log_2022}   

\usepackage{booktabs}
\usepackage{multirow}
\usepackage{amsfonts}
\usepackage{amssymb}
\usepackage{graphicx}
\usepackage[numbers,compress,sort]{natbib}

\selectfont

\makeatletter
\renewcommand{\paragraph}{%
  \@startsection{paragraph}{4}{\z@}%
                {0pt \@plus 1pt \@minus 1pt}%
                {-1em}%
                {\normalsize\bfseries\@adddotafter}%
}
\renewcommand{\section}{%
  \@startsection{section}{1}{\z@}%
                {5pt \@plus 2pt \@minus 2pt}%
                {2pt \@plus 1pt \@minus 1pt}%
                {\large\bfseries\raggedright}%
}
\makeatother

\makeatletter
\renewcommand{\@notice}{\enlargethispage{2\baselineskip}}
\makeatother
\title[MIRAGE: Measuring Interpolation and Redundancy]{MIRAGE: \textbf{M}easuring
\textbf{I}nterpolation and \textbf{R}edundancy in \textbf{A}ffinity \textbf{GE}neralization}

\providecommand{\institute}[1]{\normalfont\small #1}

\author[M. Yazdani-Jahromi et al.]{%
Mehdi Yazdani-Jahromi\\
\institute{DeepBio Scientific, Winter Park, FL}\\
\email{yazdani@deepbioscientific.com}\And
Sanjay Padhi\\
\institute{DeepBio Scientific, Winter Park, FL}\\
\email{sanjay@deepbioscientific.com}\And
Ivan Garibay\\
\institute{University of Central Florida, Orlando, FL\\ DeepBio Scientific, Winter Park, FL}\\
\email{ivan@deepbioscientific.com}
}

\begin{document}
\maketitle

\begin{abstract}
Deep-learning models now play a central role across structure-based drug design, from predicting
protein--ligand complexes and binding affinity to ligand ranking and pose generation. Recent
co-folding models have been reported to approach free-energy perturbation accuracy at substantially
lower computational cost. Yet standard evaluation based on a single held-out correlation or pooled
pose-success rate cannot distinguish transferable binding principles from repeated exposure to
related protein families in public structural databases. This distinction matters because practical
success depends on performance on genuinely novel targets.

We introduce MIRAGE, \emph{\textbf{M}easuring \textbf{I}nterpolation and \textbf{R}edundancy in
\textbf{A}ffinity \textbf{GE}neralization}, a plug-in benchmark that treats \emph{historical public family support} (through 2019) as an explicit
experimental variable. MIRAGE applies a common family-support axis to
affinity and pose prediction using matched strata, family-disjoint controls, ligand-only baselines,
and temporal evaluation.

Across affinity benchmarks, co-folder accuracy rises sharply with historical public family support,
whereas
shallow controls that cannot exploit the test family remain approximately flat. We define this
difference as the \emph{family generalization gap}. The gap is large for both co-folders and
indistinguishable from zero for every family-disjoint and trivial control. For Nesso-1 it persists
under covariate adjustment, exact-target conditioning, family balancing and clustering at 20--50\%
identity, while Boltz-2's endpoint is limited by its coverage. It localizes to the axis of historical public family support rather than to ligand chemistry, and approaches a level reproduced by family
identity alone. Rankings reverse on novel families, where a family-disjoint random forest leads both co-folders,
significantly so against Nesso-1. On one external low-support target, neither co-folder convincingly exceeds molecular weight, which
also explains a substantial fraction of performance on the standard benchmark; this is corroborative,
not population-level, evidence.  Extending the analysis beyond co-folders, gnina exhibits significant family-support dependence in
affinity rescoring whereas smina does not; in pose prediction, learned engines without MSA input have larger point-estimate gaps than the smina
redocking reference.

These results identify \emph{redundancy-driven inflation} as distinct from conventional leakage. We
propose reporting learned methods as performance across protein-family support, together with their
excess over an empirical baseline that does not improve with family support. We release MIRAGE as an installable benchmark and dataset
for affinity and pose generalization.
\end{abstract}

\section{Introduction}

Deep learning now occupies every step of structure-based drug design: co-folding
models (AlphaFold3 \citep{abramson2024alphafold3}, Boltz-1/2
\citep{wohlwend2024boltz1,passaro2025boltz2}, Chai-1 \citep{chai2024chai1}, Nesso-1
\citep{valencelabs2026nesso1}, TerraBind \citep{terray2026terrabind}) predict affinity directly, with
Boltz-2 reported to approach free-energy-perturbation accuracy; learned scoring functions such as gnina
\citep{mcnutt2021gnina} rescore poses; and neural dockers such as DiffDock-L \citep{corso2024diffdockl}
and SigmaDock \citep{prat2026sigmadock} generate them. Each is summarised by one held-out number, either a correlation or a pooled
pose-success rate, which cannot separate a model that has learned binding physics from one
recognising protein families it has seen many times in the PDB. The distinction is decisive: a programme succeeds or fails on \emph{novel} targets, least represented
in public data.

Prior work shows that ligand-based and docking benchmarks reward memorisation
\citep{wallach2018memorization,chen2019dude}, that affinity networks rely on train--test similarity
rather than physics \citep{volkov2022frustration,kanakala2023latentbiases}, and that de-leaking the
corpus lowers apparent accuracy \citep{li2023leakproof,graber2025cleansplit}. What is missing is a model-agnostic account of how deep-learning accuracy, in affinity
\emph{and} in pose, depends on \emph{protein-family support}.
Runs N' Poses \citep{runsnposes2026} studies pose recovery against training similarity, NTAB \citep{ntab2026} partitions test data into \emph{ligand}-novelty tiers, and Leak-Proof PDBBind
\citep{li2023leakproof} builds protected splits and \emph{retrains} scoring functions; MIRAGE\footnote{Harness, dataset and released predictions:
\url{https://github.com/DeepBio-Scientific/MIRAGE}; see Appendices~\ref{app:discussion}
and~\ref{app:repro}.} audits \emph{frozen} frontier models across the
full family-support curve and adds an external temporal evaluation. Nesso-1's report explicitly lacks a disclosed target sequence or structure similarity analysis
\citep{valencelabs2026nesso1}, which MIRAGE supplies; Appendix~\ref{app:related} extends this
positioning.

\section{The MIRAGE benchmark}
\label{sec:design}

\paragraph{Family support and the gap.} For a target in family $f$, the public family support $S_f$ is
the \emph{inclusive} number of PDBbind \citep{liu2017pdbbind} structures in the same MMseqs2
\citep{steinegger2017mmseqs2} 30\% sequence-identity cluster, the evaluation complex included, so a
singleton family has $S_f=1$ and the support excluding the query is $S_f-1$. $S_f$ is historical public support through 2019: every evaluation complex is a 1982--2019 PDBbind
deposition (Appendix~\ref{app:setup}). We bin $S_f$ into $\{1,2\text{--}5,6\text{--}20,21\text{--}80,81\text{--}300,301+\}$ and within each bin
compute Pearson $r$ over a \emph{matched} $\mathrm{p}K$ window $[4.5,8.0]$. The \emph{family
generalization gap} $G_m=r_m(S_f{\ge}301)-r_m(S_f{=}1)$ is estimated by a two-level bootstrap over
families then ligands; the estimator, the matched window and what each endpoint measures are set out in
Appendix~\ref{app:defs}.

\paragraph{Controls and data.} Family-disjoint models provide a negative control for direct family
recognition because their fitting procedure explicitly excludes the evaluated family: a random-forest QSAR model (ECFP4
plus amino-acid composition) and a ligand-$k$NN baseline under 5-fold \emph{family-disjoint}
cross-validation, plus family-mean, molecular weight and clogp. The redundancy set is $18{,}759$ PDBbind-derived complexes with exact $K_d/K_i/$IC$_{50}$ labels,
annotated with $S_f$ and ligand nearest-neighbour Tanimoto and stratified by support $\times$ ligand
similarity so the two memorisation channels separate; the temporal set is $649$ compounds against one
low-support target with a newly released ligand and affinity series \citep{openbind2026dataset}.
Boltz-2 and Nesso-1 run from public weights at default settings, smina
\citep{koes2013smina,trott2010vina} and gnina \citep{mcnutt2021gnina} supply classical empirical and
learned docking scores, and Chai-1 and ESMFold2 ipTM enter as labelled proxies. Datasets, models and
the reporting protocol are detailed in Appendices~\ref{app:data} and~\ref{app:setup}.

\section{Results}

\begin{figure}[tb]
\centering
\includegraphics[width=0.325\linewidth]{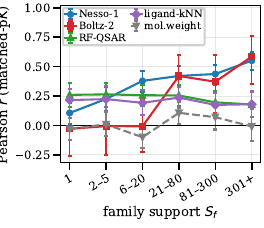}\hfill
\includegraphics[width=0.325\linewidth]{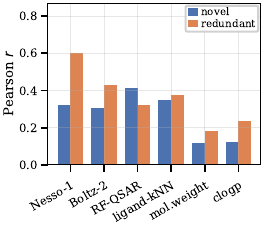}\hfill
\includegraphics[width=0.325\linewidth]{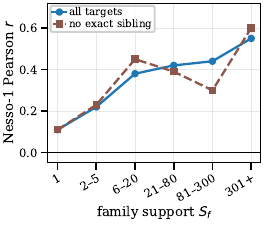}
\caption{\textbf{Left:} matched-$\mathrm{p}K$ accuracy versus family support (bootstrap 95\% CIs);
co-folders rise, family-disjoint controls stay flat. \textbf{Centre:} the co-folders' advantage is
confined to redundant families; on novel families the controls lead. \textbf{Right:} the gradient is
not exact-target repetition: restricting to targets whose sequence appears nowhere else in the corpus
leaves Nesso-1's curve intact ($G_m=+0.49$ against $+0.45$ unrestricted; Appendix~\ref{app:exposure}).}
\label{fig:dose}
\end{figure}

\paragraph{Redundancy-driven inflation: co-folder accuracy tracks family support, control accuracy does
not.} Table~\ref{tab:gap} and Figure~\ref{fig:dose} report the central finding, and
Appendix~\ref{app:dose} the full curve. Because prediction coverage differs by method, every control
is scored on the co-folder's own complexes and $\Delta G$ comes from a shared two-level bootstrap. On
Nesso-1's $2{,}147$ matched complexes the gap is $+0.446$ against $-0.082$ for RF-QSAR, $-0.035$ for
ligand-$k$NN, $+0.067$ for gnina and $+0.130$ for smina; the paired difference against RF-QSAR is
$+0.528$ $[+0.26,+0.75]$ and every paired contrast excludes zero. On Boltz-2's $373$ matched complexes
the pattern is the same and larger, with wider intervals and the contrast against gnina spanning
zero. The two behave differently under stress: Nesso-1's gap stays
significant under family balancing, assay restriction and clustering at 20--50\% identity, whereas
Boltz-2 keeps a significant continuous support slope but its family-balanced endpoint interval spans
zero, a consequence of its $583$-complex coverage rather than a reversal of direction
(Appendices~\ref{app:data}, \ref{app:dose} and~\ref{app:discussion}). The controls are not weaker co-folders; they respond \emph{differently} to familiarity. The gap is
robust to measurement type for Nesso-1 ($K_d$ only $+0.50$, $K_d{+}K_i$ $+0.49$); Boltz-2's per-assay
strata are underpowered at its current coverage (Appendix~\ref{app:data}). The matched-$\mathrm{p}K$ strata remove the major correlation advantage produced by wider label ranges
in well-populated families, and the controls' flatness is the expected signature of a model with nothing to recognise. We use
\emph{inflation}, not leakage: proprietary training sets cannot be reconstructed. 

\begin{table}[tb]
\centering\footnotesize
\caption{Family generalization gap $G_m=r(S_f{\ge}301)-r(S_f{=}1)$ with every method scored on the
\emph{same} complexes as the co-folder it is compared against, since prediction coverage differs by
method. $\Delta G$ is the paired difference from a shared two-level bootstrap. Restricting to matched
samples moves the controls' gaps more negative than the full-sample values, so an unpaired comparison
understates the interaction. The corresponding panel on Boltz-2's $373$ complexes is
Table~\ref{tab:gapboltz}.}
\label{tab:gap}
\setlength{\tabcolsep}{6pt}
\begin{tabular}{lccc}
\toprule
Method & $G_m$ & $\Delta G$ vs co-folder & 95\% CI \\
\midrule
\multicolumn{4}{l}{\emph{on Nesso-1's $2{,}147$ matched complexes}} \\
Nesso-1      & $\mathbf{+0.446}$ & -- & -- \\
RF-QSAR      & $-0.082$ & $\mathbf{+0.528}$ & $[+0.26,+0.75]$ \\
ligand-$k$NN & $-0.035$ & $+0.481$ & $[+0.25,+0.72]$ \\
gnina        & $+0.067$ & $+0.379$ & $[+0.17,+0.59]$ \\
smina        & $+0.130$ & $+0.318$ & $[+0.09,+0.59]$ \\
\bottomrule
\end{tabular}
\end{table}

\paragraph{The dominant measured axis is protein-family support, consistent with protein-family
familiarity, and the effect is not explained by the measured covariates.} In a per-target regression of absolute error on $\log_{10}S_f$, controlling for ligand similarity,
protein length, publication year, within-family affinity variance, assay type and affinity regime, the
family-support coefficient stays large and significant for Nesso-1 ($-0.130$, $t=-5.1$) and Boltz-2
($-0.152$, $t=-3.4$). Standard errors are CR1
cluster-robust, clustered on protein family ($1{,}410$ and $373$ clusters respectively), matching the
dependence structure the two-level bootstrap assumes elsewhere. Joint stratification gives the same pattern: Nesso-1 rises from about $0.1$ in the lowest support
stratum to about $0.5$ in the highest, with no ligand-similarity gradient in the memorisation
direction. No single control is sufficient; the claim rests on their conjunction: family-disjoint negative
controls, family-clustered covariate adjustment, exact-repetition decomposition, family balancing and
leave-one-family-out. Together they identify protein-family support as the primary measured
memorisation channel, complementary to the ligand-similarity channel of ligand-tiered benchmarks
\citep{ntab2026} (Appendix~\ref{app:2d}). The gradient is not explained by exact-target repetition. Adding a repetition indicator leaves the
coefficient unchanged (Nesso-1 $-0.130\to-0.135$, $t=-5.2$) with the indicator insignificant, and
restricting to targets whose sequence appears nowhere else in the corpus leaves $G_m$ undiminished
($+0.49$ $[+0.23,+0.74]$ versus $+0.45$) with every control flat. The gradient is steeper among
repeated targets than unique ones but clearly present in both, so repetition amplifies rather than
creates it (Appendix~\ref{app:exposure}).

\paragraph{Simple diagnostic baselines show how much this channel alone can explain.} A family-mean
predictor reaches $r=0.564$ under random splitting, close to the pooled co-folder $r=0.588$, and a
protein-only forest, with no access to the ligand and so no basis for pair-specific affinity, reaches
$0.637$ to $0.653$ under random splitting but $0.248$ to $0.330$ when families are held out
(Appendix~\ref{app:modality}). Much of what a random split measures is recoverable protein-family
identity rather than transferable modelling of protein--ligand binding.

\paragraph{On novel families the ranking inverts, and an external low-support target agrees.} On novel families ($S_f\le5$) the family-disjoint random forest leads both co-folders, each contrast
paired on the complexes the two methods share. Against Nesso-1 it scores $0.411$ to $0.324$,
$\Delta=+0.087$ $[+0.007,+0.167]$, excluding zero; against Boltz-2, on its $189$-complex coverage,
$0.505$ to $0.308$, $\Delta=+0.197$ $[-0.018,+0.405]$, the highest point estimate but not separated
from zero. Ligand-$k$NN is indistinguishable from either. A shallow model with no structural input therefore matches or exceeds the evaluated co-folders on
low-support families, the regime that defines a new programme. The external temporal arm asks whether the low-support result reproduces on a newly released ligand
series: the target has two prior public family structures, counted against RCSB rather than the
PDBbind reference corpus (Appendix~\ref{app:temporal}), and its ligands post-date every documented
cutoff even though the protein fold does not (Appendix~\ref{app:setup}). Across $643$ compounds assayed
against this target, within-target ranking performance was modest throughout. Molecular weight alone
reached a Spearman correlation of $\rho=0.469$. Nesso-1 gave the highest value ($\rho=0.486$), a margin
that its own report describes as not statistically convincing \citep{valencelabs2026nesso1}, while
Boltz-2 ($\rho=0.395$), gnina, smina, AEV-PLIG \citep{valsson2025aevplig} and AQ-Affinity all fell
below the molecular-weight baseline, which the OpenBind release likewise identifies as strong for this
campaign \citep{openbind2026dataset}. We note that this is a single external target rather than population-level evidence (Appendix~\ref{app:temporal}).

\paragraph{The contrast extends beyond co-folders: gnina depends on family support, smina does not.}
Re-scoring the crystal pose separates two regimes. smina is weak overall ($r=0.095$) and does not
improve with family support, its slope having the \emph{opposite} sign to the learned models'
($t=+2.4$); gnina's CNN is far stronger ($r=0.444$) and carries the co-folders' dependence ($t=-2.9$),
with its skill concentrated on families in its PDBbind training set (Appendix~\ref{app:docking}). 

\paragraph{For Boltz-2, an MSA rescues novel families while leaving familiar ones unchanged.} Boltz-2 provides the cleanest controlled intervention: with weights, architecture and target set fixed,
a ColabFold MSA raises novel-family success from $2\%$ to $46\%$ while redundant success stays at
exactly $66\%$, collapsing pose-$G_m$ from $+0.64$ to $+0.20$. The effect is one-directional, flipping
$22/50$ novel complexes to success and \emph{none} the reverse, and both settings posed all $100$
targets. Chai-1 gains on novel families too ($22\%$ to $33\%$) but with substantial run attrition, so
we report only its novel-side gain and treat it as partial directional replication
(Appendix~\ref{app:pose}). Supplying evolutionary context removes much of the same pose deficit associated with low public family
support, so that deficit is \emph{localisable} and is at least in part a missing-input problem.

\paragraph{Pose prediction.} We redock crystal ligands into $50$ singleton and $50$ redundant targets
and score symmetry-corrected RMSD, with the protocol, engines and full results in
Appendix~\ref{app:pose}. smina itself shows a $+0.23$ pose gap, from $62\%$ success on novel targets to
$86\%$ on redundant ones (Table~\ref{tab:pose}).
Because smina does not learn from the evaluated family during this experiment, we use this as a
reference for differences in redocking difficulty between the two target sets. Deep dockers sit differently above it. SigmaDock adds an excess of only $\approx0.05$
($80\%\to52\%$); DiffDock-L adds $+0.13$ and collapses to $18\%$ on novel families (median RMSD
$14$\,\AA) while holding $54\%$ on redundant ones; single-sequence co-folders are most extreme, Boltz-2 reaching $+0.64$ with $2\%$ novel success against
$66\%$ redundant. A pooled success rate cannot tell these apart; pose-$G_m$ does. The $+0.23$ difficulty reference is like-for-like only for engines given the same crystal receptor, which is the case for
SigmaDock and DiffDock-L; the co-folders build the complex de novo and receive strictly less
information, so their gaps are not the reference plus a memorisation excess and we do not subtract. For
the co-folders the controlled comparison is the within-model MSA intervention above, where
architecture, receptor information and target set are all held fixed.

\paragraph{The missing control on the field's standard generalization benchmark.} ATOM3D LBA \citep{townshend2021atom3d}, PDBbind-refined complexes split at 30\% sequence identity, is
the standard structure-based affinity-generalization benchmark, and recent
geometric networks report large gains on it, IPBind \citep{li2025ipbind} reaching Pearson $0.732$. Yet no published LBA30 result reports a ligand-only or molecular-weight control. We supply one
(Appendix~\ref{app:lba}). IPBind exceeds it by $\approx0.29$, so the top methods do learn structural
signal; but molecular weight \emph{alone} ($0.436$ on the full $490$-entry split) matches or exceeds
much of the leaderboard, so small margins over those methods were never contextualised against a
trivial descriptor. Ligand-only reaches $0.440$ here but $0.31$--$0.41$ on MIRAGE novel families:
different stringencies, and only one resembles a new target.

\section{Discussion}

\paragraph{What we claim.} On this PDBbind-derived evaluation, Nesso-1's accuracy is strongly
associated with PDBbind family density, and baselines scored on the same complexes show significantly
less of that dependence ($\Delta G=+0.53$ $[+0.26,+0.75]$), under covariate adjustment with
family-clustered errors, exact-target conditioning, family balancing, leave-one-family-out and 20--50\%
clustering. Family density remains a proxy rather than evidence of training exposure, and the design
does not separate memorisation from other family-correlated difficulty.

\paragraph{What this buys a discovery programme.} MIRAGE turns a benchmark number into a decision rule: before trusting a learned scorer or docker, compute the target's family support; for the evaluated
models, the low-support values of roughly $0.1$--$0.3$ are more relevant to a new target than their
pooled correlations. Selecting on pooled accuracy selects
for familiarity with families already solved.

\paragraph{What this means for people building models.} Validate on family-disjoint splits \emph{during} development, not only
at the end: a protein-only model reaches $r\approx0.64$ under random splitting, so architectures,
hyperparameters and checkpoints chosen on a random split are partly chosen for recognition. Along the observed support curve the gains are concentrated in already-dense families rather than at
the novel end, and supplying missing evolutionary context is the one intervention we test directly. And the dependence is not confined to a trained affinity head, since Chai-1's ipTM carries it with no
affinity head at all, so changing the affinity readout alone cannot be assumed to remove it.
SigmaDock, within $\approx0.05$ of the empirical reference, shows the gap is not a law.

\paragraph{A family-insensitive empirical scorer provides a useful reference.} The classical empirical scorer is weaker overall but does not become more accurate as public family
support increases. MIRAGE quantifies this support dependence through $G_m$. smina is an empirical scoring function, not a free-energy
calculation, and two scorers do not establish a law, so we offer it as a reference point rather than
as a general transfer benchmark. What our experiments license is narrower than a prescription: the
observed gains are concentrated in structurally supported families, while supplying evolutionary
context substantially improves novel-family pose recovery. Whether changing the composition of
training data, or adding physically grounded information, produces the same effect remains to be
tested. In the meantime, reporting a learned method as its \emph{excess over a family-insensitive
empirical reference} costs nothing and says more than a pooled correlation. A model that beats the empirical baseline on familiar families and merely matches it on unfamiliar
ones has not earned a prospective role; one that beats it on singletons has. 

{\small
\bibliographystyle{unsrtnat}
\bibliography{references}
}

\appendix

\section*{Appendix}

In this appendix we provide an extended related-work and positioning discussion
(Appendix~\ref{app:related}), the definitions and estimation procedure (\ref{app:defs}), dataset
construction and the reporting protocol (\ref{app:data}), the experimental setup (\ref{app:setup}),
the full affinity dose--response with covariate adjustment (\ref{app:dose}), the two-dimensional localisation of the effect (\ref{app:2d}), the test of exact-target repetition
against family familiarity (\ref{app:exposure}), the family-identity baseline and modality ablation
(\ref{app:modality}), the external low-support temporal evaluation (\ref{app:temporal}), the
confidence-proxy co-folders (\ref{app:proxy}), the docking scoring functions (\ref{app:docking}), the
pose MSA ablation and hardware notes (\ref{app:pose}), the family-balanced sensitivity analysis (\ref{app:balanced}), the LBA30 control (\ref{app:lba}), an
extended discussion with full limitations (\ref{app:discussion}), and the released artifact
(\ref{app:repro}).

\section{Extended related work and positioning}
\label{app:related}

\paragraph{Co-folding models that predict affinity.} AlphaFold2 \citep{jumper2021alphafold2} was
extended to protein--ligand complexes by AlphaFold3 \citep{abramson2024alphafold3} and by the open
reimplementations Boltz-1 \citep{wohlwend2024boltz1}, Chai-1 \citep{chai2024chai1} and
RoseTTAFold-All-Atom \citep{krishna2024rfaa}. Boltz-2 \citep{passaro2025boltz2}, Nesso-1
\citep{valencelabs2026nesso1} and TerraBind \citep{terray2026terrabind} go further and predict
affinity directly, and each is evaluated by an overall correlation on a held-out set. That reporting
convention is the object of this paper: it is not wrong, but it is not decomposable, and the
decomposition turns out to matter.

\paragraph{Scoring functions and QSAR baselines.} Empirical scoring functions (Vina
\citep{trott2010vina}, smina \citep{koes2013smina}) and learned ones (RF-Score
\citep{ballester2010rfscore}, Gnina \citep{mcnutt2021gnina}, AEV-PLIG \citep{valsson2025aevplig}) are
trained and tested on PDBbind \citep{liu2017pdbbind} and CASF-2016 \citep{su2019casf}, with labels
drawn from BindingDB \citep{gilson2016bindingdb} and ChEMBL \citep{mendez2019chembl} and accuracy
judged against FEP+ \citep{wang2015fep}. These share a training corpus with the co-folders, which is
why they inherit the same family structure (Appendix~\ref{app:docking}).

\paragraph{Memorisation and benchmark critiques.} Wallach and Heifets
\citep{wallach2018memorization} and Chen et al.\ \citep{chen2019dude} traced ligand-based and docking
``success'' to dataset bias rather than modelled physics. Volkov et al.\ \citep{volkov2022frustration}
found that ligand-only and protein-only models match full-complex models and that nearest-neighbour
baselines are already strong, a result our modality ablation reproduces on the family axis
(Appendix~\ref{app:modality}). Kanakala et al.\ \citep{kanakala2023latentbiases} documented latent
biases in affinity datasets, and MISATO \citep{siebenmorgen2024misato} re-curated the underlying data.
MIRAGE differs from this line in what it does with the observation: rather than arguing that
benchmarks are contaminated, it makes redundancy an axis and measures the slope along it.

\paragraph{Positioning against recent split-aware benchmarks.} Three recent efforts are the closest
neighbours, and MIRAGE is complementary to each. Runs N' Poses \citep{runsnposes2026} studies pose
recovery as a function of training-set similarity; MIRAGE studies affinity \emph{and} pose as a
function of protein-family support, and adds an empirical-baseline reference. NTAB, the Novelty-Tiered Affinity Benchmark \citep{ntab2026},
partitions test data into \emph{ligand}-novelty tiers to ask whether models learn biophysics or
memorise training patterns; MIRAGE makes protein-family familiarity the primary axis while
controlling ligand novelty, so the two benchmarks close different memorisation
channels and are best used together. Leak-Proof PDBBind \citep{li2023leakproof} constructs protected
splits and \emph{retrains} conventional scoring functions, and CleanSplit
\citep{graber2025cleansplit} shows that resolving data bias changes measured generalization; MIRAGE
instead audits \emph{frozen} frontier models across the full support curve, which is the setting a
practitioner actually faces when choosing an off-the-shelf model. Sheridan
\citep{sheridan2013timesplit} motivates the time-split view our external temporal arm adopts. Nesso-1's own
report \citep{valencelabs2026nesso1} presents chemical-similarity analyses but explicitly lacks a
disclosed target sequence or structure similarity analysis; MIRAGE supplies precisely that missing
evaluation, on public weights.

\section{Definitions and estimation}
\label{app:defs}

\paragraph{Family support.} For a target in family $f$, the \emph{PDBbind family support} $S_f$ is the
\emph{inclusive} number of PDBbind \citep{liu2017pdbbind} structures falling in the same MMseqs2
\citep{steinegger2017mmseqs2} 30\% sequence-identity cluster, counting the evaluation complex itself.
A singleton family therefore has $S_f=1$, and the number of \emph{other} supporting structures is
$S_f-1$; $S_f$ is the quantity plotted, binned and used in every equation in this paper, and it is the
released dataset column \texttt{family\_size}. $S_f$ is therefore a PDBbind research-density measure through 2019, not a per-model training count. It
requires both a deposited structure \emph{and} an affinity label, so it conflates structural
representation, label availability, target popularity, crystallisability and medicinal-chemistry
investment; we use it as a proxy for those jointly and say so wherever it is interpreted: every
evaluation complex is a 1982--2019 PDBbind deposition. We reserve ``pre-cutoff'' for statements about
a model whose cutoff is actually documented (Table~\ref{tab:cutoffs}); for Nesso-1 and ESMFold2, whose
cutoffs are undisclosed, we make no such statement, and DiffDock-L's 2018 split predates part of the
corpus, which we note rather than gloss. It is a
support proxy, not a training count: we cannot enumerate proprietary training sets, and the whole
argument is built so that we do not have to. We bin $S_f$ into $\{1,\,2\text{--}5,\,6\text{--}20,\,21\text{--}80,\,81\text{--}300,\,301{+}\}$,
which is roughly logarithmic and keeps each bin populated. Clustering uses MMseqs2 \texttt{easy-cluster}
at 30\% sequence identity with 90\% bidirectional coverage; for multichain entries the longest
polypeptide chain defines the target sequence; duplicate PDB identifiers are removed before
clustering, and error statistics are computed per complex, with per-target aggregates stated as such
wherever they are used. Every reported $r$ is accompanied by its complex count and, where it bears on
inference, its family count.

\paragraph{What $S_f$ does and does not separate.} $S_f$ conflates four kinds of exposure, which we
separate explicitly. (i) \emph{Exact-complex inclusion in the MIRAGE corpus} is constant: every evaluation complex
contributes one structure to its own inclusive family count, and because this contribution is
identical across support strata it cannot generate the support gradient. This is a statement about MIRAGE's construction and requires no knowledge of proprietary training
data. Correspondingly, the exact-target conditioning of Appendix~\ref{app:exposure} excludes repeats
\emph{inside PDBbind}, not repeats inside any model's training data, which we cannot observe. (ii) \emph{Exact-target exposure}, an identical protein sequence appearing elsewhere in the corpus
with a different ligand, is absent by construction in the $S_f=1$ bin and reaches $0.74$ of complexes
in the $301{+}$ bin (Table~\ref{tab:exposure}). (iii) \emph{Close-homolog support} and (iv) \emph{remote family support}
are what the 30\% clustering aggregates. $S_f$ therefore conflates exact-target repetition with homolog and remote-family support, and these
are strongly correlated in PDBbind. We separate them empirically (Appendix~\ref{app:exposure}): the
gradient is carried by family support, it persists under conditioning on exact repetition, and it is
undiminished among targets with no identical-sequence sibling anywhere in the corpus. What we do not
claim is exposure outside this corpus: a target unique here may still repeat in a model's private
training data.

\begin{table}[!ht]
\centering\small
\caption{Exposure decomposition across support bins. Exact-complex exposure is held at ``present'' in
every bin; exact-target exposure is what varies with $S_f$ and is disclosed here.}
\label{tab:exposure}
\begin{tabular}{lccc}
\toprule
$S_f$ bin & $n$ & median other support ($S_f-1$) & fraction with an identical-sequence sibling \\
\midrule
1       & 890  & 0   & 0.000 \\
2--5    & 2340 & 2   & 0.379 \\
6--20   & 3977 & 11  & 0.604 \\
21--80  & 4685 & 41  & 0.707 \\
81--300 & 2897 & 127 & 0.686 \\
301+    & 3970 & 732 & 0.739 \\
\bottomrule
\end{tabular}
\end{table}

\paragraph{Matched strata, and why they are necessary.} Pearson $r$ depends on the spread of the
labels being correlated. Families that are heavily represented in the PDB also tend to have wider
affinity ranges, so a naive per-bin correlation would rise with support even for a model with no
family knowledge at all. We therefore compute $r$ within a fixed $\mathrm{p}K$ window $[4.5,8.0]$ inside each bin, holding
label spread, and hence attainable correlation, approximately constant. To be explicit about the
estimator: for bin $b$ with complex set $\mathcal{C}_b$ (all complexes in that bin whose label falls
in the window),
\begin{equation}
r_m(b)\;=\;\frac{\sum_{i\in\mathcal{C}_b}(\hat{y}_i-\bar{\hat y}_b)(y_i-\bar y_b)}
{\sqrt{\sum_{i\in\mathcal{C}_b}(\hat{y}_i-\bar{\hat y}_b)^2}\;
\sqrt{\sum_{i\in\mathcal{C}_b}(y_i-\bar y_b)^2}},
\end{equation}
an ordinary per-complex Pearson correlation over the bin, not a family-weighted one and not an average
of per-family correlations. A per-family correlation is undefined in the singleton bin, where each
family contributes one complex, which is why we do not use one. Family-level dependence enters the interval through the two-level bootstrap, which resamples families
first and ligands within them, and every regression uses CR1 errors clustered on family. Weighting,
however, changes the \emph{estimand} rather than the interval: a complex-weighted $r$ estimates
performance for a random complex drawn from the bin, not for a random family. Because the $301{+}$ bin
draws thousands of complexes from seven families, we report a family-balanced estimator alongside it
(Appendix~\ref{app:balanced}). The fixed window controls one specific and serious source of inflation, namely correlation rising with
label spread, but it does not by itself exclude every form of differential difficulty across strata.
That work is done jointly by the family-disjoint controls, the covariate regression with
family-clustered errors, the exact-target decomposition and the family-balanced estimand. Throughout, \emph{pooled} numbers are computed over all
complexes at once and \emph{matched} numbers within this window inside each support bin; the two can
differ substantially, and every claim about the family-support gradient uses matched numbers.

\paragraph{The endpoint does not rest on one high-support family.} Deleting each of the seven
families in the $S_f\ge301$ cell in turn and recomputing $G_m$ against the unchanged singleton bin
changes Nesso-1's gap by at most $0.015$ (range $+0.431$ to $+0.456$; $+0.445$ when the largest,
$24.7\%$ of the cell, is removed), so the endpoint contrast is not carried by one heavily sampled
family (Table~\ref{tab:loo}). Every control stays within $\pm0.08$ of zero under every deletion.
Per-family detail, including family id, complexes dropped and share of the cell, is in
\texttt{results/sensitivity\_stats.json} under \texttt{leave\_one\_family\_out}.

\begin{table}[!ht]
\centering\footnotesize
\caption{Leave-one-high-support-family-out. Each of the seven families in the $S_f\ge301$ cell is
dropped in turn and $G_m$ recomputed against the unchanged singleton bin. Values are quoted to three
decimals here because the quantity of interest is the size of the movement; $G_m$ is quoted to two
decimals everywhere else, so Nesso-1's $+0.446$ is the $+0.45$ of Table~\ref{tab:gap} and the $+0.45$
complex-weighted column of Table~\ref{tab:balanced}.}
\label{tab:loo}
\begin{tabular}{lccl}
\toprule
Method & full $G_m$ & leave-one-out range & dropping the largest family \\
\midrule
Nesso-1      & $+0.446$ & $\mathbf{[+0.431,+0.456]}$ & 3i4b, 83 complexes, $24.7\%$ of the cell $\to$ $+0.445$ \\
Boltz-2      & $+0.615$ & $[+0.559,+0.702]$ & 4feq, 16 complexes, $25.8\%$ $\to$ $+0.559$ \\
gnina        & $+0.068$ & $[+0.034,+0.094]$ & 3i4b, $24.7\%$ $\to$ $+0.094$ \\
smina        & $-0.036$ & $[-0.039,+0.078]$ & 3i4b, $22.1\%$ $\to$ $+0.078$ \\
RF-QSAR      & $-0.012$ & $[-0.047,+0.052]$ & 3i4b, $22.1\%$ $\to$ $+0.002$ \\
ligand-$k$NN & $-0.025$ & $[-0.060,+0.020]$ & 3i4b, $22.1\%$ $\to$ $-0.026$ \\
\bottomrule
\end{tabular}
\end{table}

\paragraph{The family generalization gap.}
\begin{equation}
G_m \;=\; r_m(S_f \ge 301) \;-\; r_m(S_f = 1),
\end{equation}
the change in matched accuracy from singleton families to well-supported ones. $G_m>0$ means accuracy
is strongly associated with public family support. $G_m$ is a property of a \emph{model on a benchmark}, not of a model
alone, and it is the quantity we argue should be reported alongside the headline correlation.

\paragraph{Two-level bootstrap.} Complexes are not independent: a single well-studied family
contributes many ligands. We therefore resample \emph{families} with replacement and then resample
\emph{ligands within each selected family}, recomputing $G_m$ on each replicate. The resulting
interval reflects both family-level and ligand-level variability, and is materially wider than a naive
per-complex bootstrap, which is why Boltz-2's interval, $[+0.12,+1.01]$, is wide despite a
large point estimate.

\paragraph{Family-disjoint controls.} Family-disjoint models provide a negative control for direct
family recognition because their fitting procedure explicitly excludes the evaluated family. We
evaluate a random-forest QSAR model (ECFP4 fingerprints concatenated with amino-acid composition) and
a ligand-$k$NN baseline under 5-fold cross-validation with \texttt{GroupKFold} on family id, so every
prediction is made for a family absent from that fold's training data. These are not meant to be strong models. They are negative controls for \emph{direct family
recognition}: a model that never sees the test family cannot recognise it, so any residual $G_m$ they
show measures whatever else covaries with family support, such as chemical tractability, pocket
canonicality or label quality. Their measured $G_m\approx0$ is therefore an empirical result, not a
guaranteed null, and it is informative precisely because it bounds those alternative explanations
rather than assuming them away.

\section{Dataset construction and reporting protocol}
\label{app:data}

\paragraph{Redundancy set.} $18{,}759$ PDBbind-derived complexes carrying an exact $K_d$, $K_i$ or
IC$_{50}$ label converted to $\mathrm{p}K$. Each complex is annotated with its protein-family support
$S_f$, its ligand nearest-neighbour ECFP4 Tanimoto to the rest of the corpus, protein length,
deposition year, measurement type, and the affinity variance of its family. A balanced $3{,}360$-target ``core'' subset (560 per support bin) is shipped for expensive models.
Nesso-1 and Boltz-2 are evaluated on that core subset and the controls on the full set, so every
comparison between them is made on matched complexes (Table~\ref{tab:gap}) and every table states its
sample.

\paragraph{Temporal set.} $649$ compounds with measured $K_d$ against a single low-support target,
the OpenBind EV-A71 2A protease \citep{openbind2026dataset}, released CC0. The protein has two prior public family structures, counted against RCSB rather than the PDBbind
reference corpus; the compound series and its measurements post-date every documented cutoff
(Table~\ref{tab:cutoffs}). Of
these, $643$ have complete predictions from every method. Molecular-weight, clogp, Gnina, smina,
AEV-PLIG and AQ-Affinity baselines ship with it. This arm exists because $S_f$ is a proxy; a low-support target whose ligands post-date the documented
cutoffs asks the question directly, at the cost of $n=1$ target and of the per-model qualifications in
Table~\ref{tab:cutoffs}.

\paragraph{Measurement-type sensitivity.} $K_d$, $K_i$ and IC$_{50}$ are not interchangeable, so we
recompute $G_m$ within measurement type (Table~\ref{tab:assay}). Nesso-1's gap is robust to every
restriction, including the $K_d$-only and $K_d{+}K_i$ subsets that avoid IC$_{50}$ entirely. Boltz-2's is not stable across the same restrictions, and its per-assay cells are very small: the
$301{+}$/$K_d$ cell contains six complexes. This is one of three coverage-driven failures discussed
together in Appendix~\ref{app:discussion}.

\begin{table}[!ht]
\centering\small
\caption{$G_m$ recomputed within measurement type (matched $\mathrm{p}K$ window, singleton versus
$301{+}$ bins). Nesso-1's gap holds under every restriction; Boltz-2's per-assay strata are
underpowered at its current coverage.}
\label{tab:assay}
\begin{tabular}{lccccc}
\toprule
Method & all & $K_d$ & $K_i$ & $K_d{+}K_i$ & IC$_{50}$ \\
\midrule
Nesso-1      & \textbf{+0.45} & $+0.50$ & $+0.61$ & \textbf{+0.49} & $+0.38$ \\
Boltz-2      & \textbf{+0.62} & $+0.80$ ($n{=}46/6$) & $-0.06$ ($n{=}12/17$) & \textbf{+0.38} & $+0.62$ ($n{=}6/39$) \\
RF-QSAR      & $-0.01$ & $+0.09$ & $-0.17$ & $-0.01$ & $-0.13$ \\
ligand-$k$NN & $-0.03$ & $+0.02$ & $-0.12$ & $-0.02$ & $+0.00$ \\
\bottomrule
\end{tabular}
\end{table}

\paragraph{Two-dimensional design.} MIRAGE stratifies by protein-family support $\times$ ligand
similarity, because the two memorisation channels are different and a benchmark that matches only
affinity distributions cannot separate them: a co-folder may exploit protein-family familiarity while
a ligand model exploits chemical similarity, and both would show up as ``accuracy''. The localisation
result in Appendix~\ref{app:2d} is what this design buys.

\paragraph{Reporting protocol.} We recommend, and the harness implements: within-stratum correlation
rather than pooled-across-target correlation, computed as in Appendix~\ref{app:defs} with family
dependence carried by the interval rather than by a weighted point estimate; Pearson $r$, Spearman $\rho$, centered RMSE and pairwise ranking accuracy reported
together; the two-level bootstrap of Appendix~\ref{app:defs}; a sensitivity check over the clustering
choice; and a mandatory baseline suite of family-mean (target identity), molecular weight, clogp,
ligand-$k$NN and family-disjoint RF-QSAR. The OpenBind score is reported as \emph{one external target}, never as population-level proof.

\section{Experimental setup}
\label{app:setup}

We evaluate five classes of method, all from public weights at default settings.

\paragraph{Affinity co-folders.} Boltz-2 \citep{passaro2025boltz2} with cuEquivariance kernels and
ColabFold MSAs, and Nesso-1 \citep{valencelabs2026nesso1} with ESM-2 embeddings
\citep{rives2021esm}. Both output $\log_{10}(\mathrm{IC}_{50})$, which we negate to a
higher-is-stronger score before correlating.

\paragraph{Confidence-proxy co-folders.} Chai-1 \citep{chai2024chai1} and ESMFold2 predict structure
and an interface confidence (ipTM) but \emph{not} affinity. We score ipTM as an affinity proxy and
label it as such throughout; it is included because it tests whether family dependence lives in the
learned structural representation or only in a trained affinity head (Appendix~\ref{app:proxy}).

\paragraph{Family-disjoint controls.} RF-QSAR (ECFP4 $+$ amino-acid composition) and ligand-$k$NN,
both under family-disjoint cross-validation as described in Appendix~\ref{app:defs}.

\paragraph{Trivial and identity baselines.} Family-mean (target identity), molecular weight, clogp.

\paragraph{Docking and pose engines.} smina \citep{koes2013smina,trott2010vina} as a classical
empirical score;
gnina CNNaffinity \citep{mcnutt2021gnina} as a learned score; and for pose, smina, SigmaDock
\citep{prat2026sigmadock}, DiffDock-L \citep{corso2024diffdockl}, Chai-1 and Boltz-2, the last two
with and without a ColabFold MSA.

AlphaFold3 \citep{abramson2024alphafold3} has open code but request-only weights and no affinity head;
MIRAGE ships it as a pluggable stub rather than an evaluated method.

\paragraph{Boltz-2 coverage.} Boltz-2 was run on the balanced core subset until compute was
exhausted, yielding $583$ of $3{,}360$ complexes. The coverage is near-uniform along the support axis,
the one dimension that could bias $G_m$: $16.8\%$, $17.0\%$, $17.5\%$, $17.7\%$, $17.3\%$ and $17.9\%$
of the six bins in ascending order. Nor is it the cheap-complexes subset a cost-truncated run would
produce: median sequence length is $266$ against $270$ for the uncovered remainder, $\mathrm{p}K$ $6.4$ against
$6.2$, and family size $22$ against $20$, none of them distinguishable (Mann--Whitney $p=0.26$, $0.30$
and $0.68$ respectively). Two variables differ weakly and we disclose them: ligand nearest-neighbour
Tanimoto and deposition year. Both have identical medians in the two groups, but a Mann--Whitney test,
which compares the whole distribution rather than the median, separates them ($p=4\times10^{-5}$ and
$p=0.03$); the difference is in distributional shape, not in central tendency. The subset is therefore a compute
limitation that is roughly uniform across the support axis rather than a biased or deliberate
selection; it is simply small, which is why Boltz-2's endpoint statistics are reported as directional.

\paragraph{Training cutoffs.} The claim that the evaluation corpus predates training is
model-specific, and structural and affinity training data have different provenance, so we state both
per model (Table~\ref{tab:cutoffs}). We assert temporal priority only where a cutoff is documented in
a primary source. Boltz-2's structural cutoff is 2023-06-01 \citep{passaro2025boltz2}, materially
later than the 2021-09-30 date used by AlphaFold3-style splits and later than the ``post-2021''
shorthand an earlier draft of this paper used; any priority claim over Boltz-2 must be checked against
2023-06-01. Separately, no evaluated affinity model discloses a cutoff for its \emph{affinity}
training data as distinct from its structures: Boltz-2's affinity head trains on ChEMBL and BindingDB
with no stated date. We therefore make no priority claim over affinity labels for any model. Neither
point threatens the redundancy arm, whose complexes are 1982--2019 PDBbind depositions and predate
every documented cutoff; both matter for the external temporal arm (Appendix~\ref{app:temporal}).

\begin{table}[!ht]
\centering\footnotesize
\caption{Structural and affinity training cutoffs per evaluated model. \emph{High} confidence means a
primary source states the date; \emph{medium} means a secondary source does and we flag it for
verification; \emph{undisclosed} means the source is explicit that no cutoff is given; \emph{to
verify} means we have not located a statement. No cutoff here is inferred from a release date. gnina, SigmaDock, AEV-PLIG and AQ-Affinity enter only
through analyses that make no temporal-priority claim, so we do not tabulate cutoffs for them.}
\label{tab:cutoffs}
\setlength{\tabcolsep}{4pt}
\resizebox{\textwidth}{!}{%
\begin{tabular}{lllll}
\toprule
Model & Structural cutoff & Affinity-data cutoff & Source & Confidence \\
\midrule
Boltz-2    & 2023-06-01 & none stated (ChEMBL/BindingDB) & report, App.~A.1 \citep{passaro2025boltz2} & high \\
Chai-1     & 2021-01-12 & n/a (no affinity head) & technical report \citep{chai2024chai1} & high \\
DiffDock-L & 2018 (PDBBind time split) & n/a (pose only) & \citep{corso2024diffdockl} & medium \\
Nesso-1    & undisclosed & undisclosed & technical report \citep{valencelabs2026nesso1} & undisclosed \\
ESMFold2   & undisclosed & n/a (no affinity head) & release materials & undisclosed \\
\bottomrule
\end{tabular}}
\end{table}

\section{Full affinity dose--response and covariate adjustment}
\label{app:dose}

Table~\ref{tab:dose} gives the complete matched-$\mathrm{p}K$ curve summarised by Figure~1 of the
body. The shape is the finding: the co-folders rise across nearly three orders of magnitude of family support (support runs from $1$
to $809$, $\log_{10}=2.91$), Boltz-2 non-monotonically, falling from $0.420$ to $0.371$ before rising
to $0.585$, while the family-disjoint controls are flat or very slightly decreasing. Boltz-2 is
worth reading closely: it is essentially uncorrelated with truth ($r\approx0$) for every family with
fewer than $21$ structures, and only becomes a useful predictor once the family is well represented.

\begin{table}[!ht]
\centering\small
\caption{Matched-$\mathrm{p}K$ Pearson $r$ by protein-family support. Co-folders rise steeply;
family-disjoint controls are flat. These are the values plotted in Figure~1 (left).}
\label{tab:dose}
\begin{tabular}{lcccccc}
\toprule
Method & \multicolumn{6}{c}{protein-family support $S_f$} \\
\cmidrule(lr){2-7}
 & 1 & 2--5 & 6--20 & 21--80 & 81--300 & 301+ \\
\midrule
Nesso-1      & 0.105 & 0.225 & 0.378 & 0.421 & 0.437 & \textbf{0.551} \\
Boltz-2      & $-0.030$ & $-0.007$ & $-0.008$ & 0.420 & 0.371 & \textbf{0.585} \\
RF-QSAR      & 0.259 & 0.264 & 0.257 & 0.254 & 0.196 & 0.177 \\
ligand-$k$NN & 0.215 & 0.223 & 0.192 & 0.236 & 0.174 & 0.180 \\
mol.\ weight & $-0.032$ & 0.009 & $-0.101$ & 0.107 & 0.070 & $-0.010$ \\
\bottomrule
\end{tabular}
\end{table}

\begin{figure}[!ht]
\centering
\includegraphics[width=0.66\linewidth]{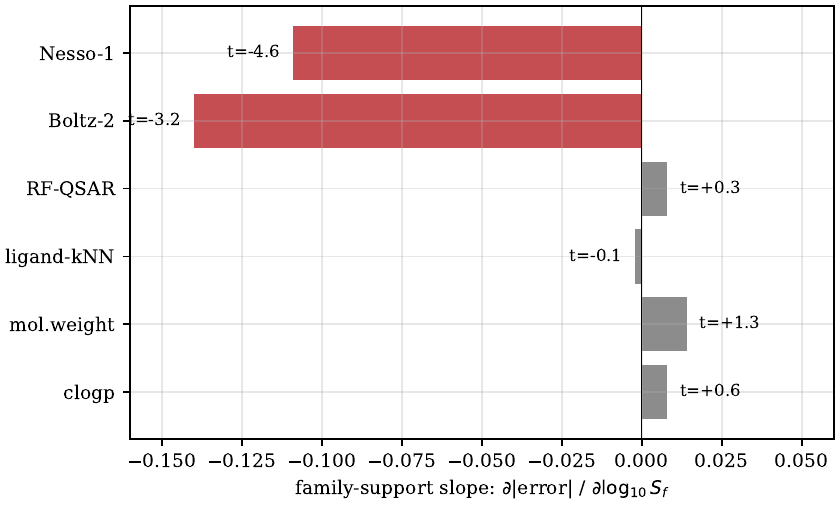}
\caption{Family-support slope $\partial|\mathrm{error}|/\partial\log_{10}S_f$ per method, with $t$
statistics (CR1 cluster-robust, clustered on protein family). Only the affinity co-folders (red) are
significantly redundancy-dependent; the controls sit at or slightly above zero.}
\label{fig:slope}
\end{figure}

\paragraph{The dependence as a single slope.} Regressing per-target absolute error on
$\log_{10}S_f$ gives one number per method, namely how much error falls per decade of family support
(Figure~\ref{fig:slope}). Only the two affinity co-folders have significantly negative slopes (Nesso $-0.109$, $t=-4.6$; Boltz
$-0.140$, $t=-3.2$). Every family-disjoint or trivial baseline is flat or slightly \emph{positive}
(RF-QSAR $+0.008$, $t=+0.3$; ligand-$k$NN $-0.002$, $t=-0.1$; molecular weight $+0.014$, $t=+1.3$;
clogp $+0.008$, $t=+0.6$). All $t$ statistics in this paper use CR1 cluster-robust standard errors
clustered on protein family; the unclustered OLS values, which the submitted version reported, are
uniformly larger in magnitude and are retained in the harness output as
\texttt{leakage\_slope\_t\_ols}. A positive slope is what a method with
no family knowledge should show, since well-supported families are, if anything, marginally harder.

\paragraph{The paired comparison on Boltz-2's coverage.} Table~\ref{tab:gapboltz} is the companion to
Table~\ref{tab:gap}, with every control scored on Boltz-2's $373$ matched complexes. The pattern is the
same and larger, with wider intervals; only the contrast against gnina spans zero.

\begin{table}[!ht]
\centering\footnotesize
\caption{Family generalization gap on Boltz-2's $373$ matched complexes, with $\Delta G$ from a shared
two-level bootstrap.}
\label{tab:gapboltz}
\begin{tabular}{lccc}
\toprule
Method & $G_m$ & $\Delta G$ vs Boltz-2 & 95\% CI \\
\midrule
Boltz-2      & $+0.615$ & -- & -- \\
RF-QSAR      & $-0.494$ & $+1.109$ & $[+0.49,+1.71]$ \\
ligand-$k$NN & $-0.304$ & $+0.919$ & $[+0.22,+1.57]$ \\
smina        & $-0.214$ & $+0.839$ & $[+0.10,+1.47]$ \\
gnina        & $+0.093$ & $+0.522$ & $[-0.08,+1.09]$ \\
\bottomrule
\end{tabular}
\end{table}

\paragraph{What each endpoint of $G_m$ measures, and the within-family gradient.} The two endpoints
of $G_m$ are not the same kind of quantity. At $S_f=1$ every family contributes one complex, so $r$ is
purely cross-target calibration; at $S_f\ge301$ it mixes between-family calibration with within-family
ligand ranking. We therefore decompose each stratum into a between-family component ($r$ of family
means) and a within-family component ($r$ of family-centred values), reported in
Table~\ref{tab:within} for $S_f\ge2$, where the within-family term exists. The gradient is present in
the within-family component, which is composition-stable and is the quantity a discovery programme
cares about: Nesso-1 rises from $-0.046$ to $+0.536$ across the strata where it is estimable, while
smina stays flat ($+0.056$ to $+0.027$). RF-QSAR is not perfectly flat here either ($+0.113$ to
$+0.299$), which we report rather than smooth over. The between-family value in the $301{+}$ stratum
rests on seven points and should not be read as a correlation.

\begin{table}[!ht]
\centering\footnotesize
\caption{Within- and between-family decomposition of the matched-$\mathrm{p}K$ correlation by support
stratum. The within-family term is undefined at $S_f=1$, where each family contributes one complex.}
\label{tab:within}
\begin{tabular}{lccccc}
\toprule
$S_f$ & Nesso-1 total & between & within & RF-QSAR within & smina within \\
\midrule
1       & $+0.105$ & $+0.105$ & undefined & undefined & undefined \\
2--5    & $+0.225$ & $+0.255$ & $-0.046$ & $+0.113$ & $+0.056$ \\
6--20   & $+0.378$ & $+0.389$ & $+0.315$ & $+0.175$ & $+0.122$ \\
21--80  & $+0.421$ & $+0.402$ & $+0.467$ & $+0.152$ & $+0.040$ \\
81--300 & $+0.437$ & $+0.558$ & $+0.446$ & $+0.195$ & $+0.070$ \\
301+    & $+0.551$ & $+0.809$ (7 pts) & $\mathbf{+0.536}$ & $+0.299$ & $+0.027$ \\
\bottomrule
\end{tabular}
\end{table}

\paragraph{Sensitivity to the correlation metric.} Recomputing $G_m$ with Spearman rather than
Pearson leaves Nesso-1 unchanged at $+0.44$ $[+0.26,+0.61]$, which closes the residual worry that the
matched window handles a Pearson-specific label-spread artifact incompletely. Boltz-2 gives $+0.57$
$[+0.03,+1.00]$, excluding zero only marginally. The controls are unchanged except smina, whose point
estimate changes sign, $-0.04$ under Pearson to $+0.09$ under Spearman, with both intervals including
zero; we report the sign change rather than quoting the more convenient value, and note that it is
consistent with our claim that smina shows none of the improvement the learned models show, and
inconsistent with describing smina as flat in either direction.

\paragraph{Sensitivity to the clustering threshold.} Family support is defined at 30\% sequence
identity, so we re-annotated the corpus with MMseqs2 at 20\%, 40\% and 50\% and recomputed $G_m$ from
the same released predictions (Table~\ref{tab:clust}). Re-clustering at 30\% reproduced the shipped
$2{,}203$ families exactly, which validates the re-annotation. Nesso-1's gap is essentially invariant
to the threshold, ranging from $+0.41$ to $+0.45$ with every interval excluding zero. The three
family-disjoint and empirical controls stay at or below zero throughout and drift slightly
\emph{negative} at 40--50\%, while gnina sits at $+0.07$ to $+0.08$ with an interval spanning zero at
every threshold. Boltz-2's point estimate falls at 40--50\% because its $S_f\ge301$ cell shrinks to
$18$ complexes there, which is also why its interval opens to roughly $\pm1$: a coverage effect rather
than a threshold effect (Appendix~\ref{app:discussion}). The corpus-level shift matters for reading
the last two columns of every row: raising the threshold from 30\% to 40\% splits the large families
and cuts redundant-bin complexes from $3{,}970$ to $1{,}248$, so those columns test a materially
smaller high-support stratum for every method. Family counts run $2{,}004$, $2{,}203$, $2{,}557$ and
$2{,}863$ across the four thresholds.

\begin{table}[!ht]
\centering\footnotesize
\caption{$G_m$ under alternative MMseqs2 identity thresholds, with two-level bootstrap 95\% CIs, and
the matched-$\mathrm{p}K$ cell sizes ($n$ at $S_f{=}1$ / $n$ at $S_f{\ge}301$) that produce them.
Nesso-1's gap is threshold-invariant; the controls stay at or below zero; Boltz-2's decline at
40--50\% tracks an $18$-complex top cell.}
\label{tab:clust}
\resizebox{\textwidth}{!}{%
\begin{tabular}{lcccc}
\toprule
Method & 20\% & 30\% (as published) & 40\% & 50\% \\
\midrule
\multicolumn{5}{l}{\emph{family generalization gap} $G_m$ [95\% CI]} \\
Nesso-1      & $\mathbf{+0.41}$ $[+0.24,+0.58]$ & $\mathbf{+0.45}$ $[+0.28,+0.61]$ & $\mathbf{+0.43}$ $[+0.12,+0.64]$ & $\mathbf{+0.42}$ $[+0.14,+0.63]$ \\
Boltz-2      & $+0.70$ $[+0.24,+1.10]$ & $+0.62$ $[+0.13,+1.02]$ & $+0.36$ $[-0.85,+1.08]$ & $+0.38$ $[-0.80,+1.11]$ \\
gnina        & $+0.08$ $[-0.11,+0.25]$ & $+0.07$ $[-0.14,+0.27]$ & $+0.08$ $[-0.22,+0.40]$ & $+0.08$ $[-0.21,+0.40]$ \\
smina        & $-0.03$ $[-0.17,+0.18]$ & $-0.04$ $[-0.18,+0.19]$ & $+0.05$ $[-0.08,+0.37]$ & $+0.05$ $[-0.09,+0.35]$ \\
RF-QSAR      & $-0.01$ $[-0.17,+0.14]$ & $-0.01$ $[-0.17,+0.16]$ & $-0.08$ $[-0.18,+0.12]$ & $-0.10$ $[-0.20,+0.11]$ \\
ligand-$k$NN & $-0.03$ $[-0.18,+0.11]$ & $-0.03$ $[-0.17,+0.11]$ & $-0.07$ $[-0.20,+0.13]$ & $-0.08$ $[-0.21,+0.12]$ \\
\midrule
\multicolumn{5}{l}{\emph{cell sizes}, $n(S_f{=}1)\,/\,n(S_f{\ge}301)$} \\
Nesso-1      & 343 / 398 & 375 / 336 & 401 / 88 & 421 / 88 \\
Boltz-2      & 59 / 70   & 64 / 62   & 70 / 18  & 74 / 18 \\
gnina        & 347 / 395 & 379 / 336 & 405 / 88 & 426 / 88 \\
smina        & 512 / 2683 & 578 / 2314 & 703 / 637 & 817 / 637 \\
RF-QSAR      & 519 / 2686 & 585 / 2317 & 711 / 638 & 830 / 638 \\
ligand-$k$NN & 519 / 2686 & 585 / 2317 & 711 / 638 & 830 / 638 \\
\bottomrule
\end{tabular}}
\end{table}

\paragraph{Adjusting for confounds.} The univariate slope invites the objection that family support
is standing in for something else. Absolute error is defined after a linear calibration $a+b\hat y$ of each method's raw score, which is
what makes $|\mathrm{error}|$ meaningful for arbitrary-scale scores such as ipTM, smina and molecular
weight. The calibration is fitted \emph{out of fold}, with \texttt{GroupKFold} folds disjoint by
family, so no family helps calibrate its own errors; refitting globally instead changes nothing
material (Nesso-1 $-0.129\to-0.130$, Boltz-2 $-0.153\to-0.152$, every other method within $0.004$), but
we adopt cross-fitting because it costs nothing and removes the objection. We regress per-target
absolute error on $\log_{10}S_f$ \emph{together with} ligand nearest-neighbour similarity, protein length, publication year,
within-family affinity variance, measurement type ($K_d$/$K_i$/IC$_{50}$), and the affinity regime
entered as $\mathrm{p}K$ and $\mathrm{p}K^2$. The family-support coefficient remains large and significant: Nesso $-0.130$ ($t=-5.1$, $1{,}410$
family clusters), Boltz $-0.152$ ($t=-3.4$, $373$ clusters), with CR1 cluster-robust standard errors
clustered on protein family. No measured covariate explains it.

\paragraph{Pooled versus matched, and why the headline number misleads.} Pooled over all complexes,
Nesso-1 scores $r=0.588$ and Boltz-2 $r=0.467$, against RF-QSAR $0.479$ and ligand-$k$NN $0.407$, a ranking that would
make the co-folders the obvious choice. Split by support, that pooled ordering
dissolves: on novel families ($S_f\le5$) the same numbers are Nesso $0.324$, Boltz $0.308$, RF-QSAR
$0.411$, ligand-$k$NN $0.350$. The pooled number is disproportionately influenced by the dense strata, which is exactly the regime a
new programme is not in; Pearson $r$ is not a weighted average of subgroup correlations, so we do not
present it as one.

\section{Localising the channel: family support $\times$ ligand similarity}
\label{app:2d}

If the co-folders were exploiting chemical similarity rather than protein familiarity, accuracy would
climb along the ligand axis. It does not (Table~\ref{tab:2d}). Reading down any column, Nesso-1's
accuracy climbs from roughly $0.1$ to roughly $0.5$ as family support increases. Reading across, the
ligand axis carries no memorisation gradient: ligand similarity is not significant in a model that
also contains family support (Tanimoto $t=+0.18$), and neither is the interaction ($t=+0.30$; CR1
clustered on family), while family support remains significant ($t=-2.59$). The one row that is not
flat runs the \emph{wrong way} for a ligand-memorisation account: within novel families, accuracy is
higher on novel ligands than on similar ones ($\Delta r=+0.26$, two-level bootstrap 95\% CI
$[+0.01,+0.50]$), not lower (Table~\ref{tab:2dcontrast}). Per-cell label spread is comparable across
that row ($\mathrm{p}K$ SD $1.02$, $0.94$, $0.88$), so attenuation does not explain it; the most likely
reading is that ``similar ligand'' cells in sparsely-sampled families are congeneric series whose
within-series ranking is the harder problem. Either way, no cell supports the hypothesis that chemical similarity is what buys co-folder accuracy.
This tests one ligand statistic, nearest-neighbour ECFP4 Tanimoto; scaffold identity, ligand size and
congeneric structure remain untested, so we claim no gradient along that statistic rather than the
exclusion of ligand chemistry. The memorisation channel for co-folders is the protein,
which is the channel that ligand-tiered benchmarks such as NTAB \citep{ntab2026} do not control.
Conversely, MIRAGE does not close the ligand channel, which is why the two designs are complementary
rather than competing.

\begin{table}[!ht]
\centering\small
\caption{Nesso-1 matched-$\mathrm{p}K$ Pearson $r$ by protein-family support $\times$ ligand
similarity, with two-level bootstrap 95\% CIs and cell $n$. The gradient runs down the family axis,
not across the ligand axis.}
\label{tab:2d}
\setlength{\tabcolsep}{4pt}
\resizebox{\textwidth}{!}{%
\begin{tabular}{lccc}
\toprule
family support $\backslash$ ligand & novel ligand & mid & similar ligand \\
\midrule
novel (1--5)    & $+0.37$ $[+0.17,+0.54]$ (170) & $+0.10$ $[-0.08,+0.28]$ (228) & $+0.11$ $[-0.05,+0.26]$ (325) \\
mid (6--80)     & $+0.44$ $[+0.25,+0.62]$ (134) & $+0.48$ $[+0.33,+0.61]$ (253) & $+0.35$ $[+0.19,+0.48]$ (353) \\
redundant (81+) & $+0.51$ $[+0.32,+0.67]$ (152) & $+0.53$ $[+0.38,+0.66]$ (270) & $+0.48$ $[+0.31,+0.64]$ (262) \\
\bottomrule
\end{tabular}}
\end{table}

\begin{table}[!ht]
\centering\small
\caption{Paired two-level bootstrap contrast within each family stratum, novel-ligand minus
similar-ligand. Only the novel-family row is significant, and its sign is opposite to what ligand
memorisation would predict.}
\label{tab:2dcontrast}
\begin{tabular}{lccc}
\toprule
family stratum & $\Delta r$ & 95\% CI & $P(\Delta>0)$ \\
\midrule
novel (1--5)    & \textbf{+0.26} & $[+0.01,+0.50]$ & 0.98 \\
mid (6--80)     & $+0.09$ & $[-0.13,+0.33]$ & 0.78 \\
redundant (81+) & $+0.03$ & $[-0.21,+0.27]$ & 0.59 \\
\bottomrule
\end{tabular}
\end{table}

\section{Exact-target repetition versus family familiarity}
\label{app:exposure}

The sharpest available objection to MIRAGE is that $S_f$ tracks \emph{the same protein appearing again
with a different ligand} rather than familiarity with a protein family: identical-sequence siblings
run from $0\%$ of the singleton bin to $73.9\%$ of the $301{+}$ bin (Table~\ref{tab:exposure}). We
measure this rather than concede it. Every complex is annotated with how many \emph{other} corpus
complexes share its protein at three stringencies, namely identical sequence, the same 99\%-identity
cluster and the same 95\%-identity cluster (MMseqs2, 90\% coverage), and we write
$E_i=\mathbf{1}\{n_{\text{exact sib}}>0\}$.

\paragraph{The family coefficient persists when exact-target exposure is added.}
Table~\ref{tab:expreg} repeats the covariate regression with $E$ included. The family-support
coefficient does not merely persist: for Nesso-1 and Boltz-2 it becomes \emph{larger}, while $E$ itself
is insignificant (Nesso-1) or positively signed (Boltz-2, where exact repetition is associated with
slightly \emph{higher} error). Replacing the indicator with continuous
$\log_{10}(1+n_{\text{exact sib}})$ gives the same answer (Nesso-1 $\beta_S=-0.124$, $t=-4.57$; sibling
term $-0.018$, $t=-0.49$). One nuance is worth reporting rather than hiding: the $\log_{10}S_f\times E$
interaction is significant for Nesso-1 ($-0.096$, $t=-2.35$), so the family gradient is steeper among
repeated targets ($-0.190$) than among unique ones ($-0.094$). It is clearly present in both.
Repetition amplifies the effect; it does not create it.

\begin{table}[!ht]
\centering\footnotesize
\caption{Covariate regression of $|\mathrm{error}|$ on $\log_{10}S_f$, with and without an
exact-repetition indicator $E$. CR1 standard errors clustered on family.}
\label{tab:expreg}
\setlength{\tabcolsep}{5pt}
\begin{tabular}{lcccccc}
\toprule
Method & $\beta_S$ alone & $t$ & $\beta_S$ given $E$ & $t$ & $\beta_E$ & $t_E$ \\
\midrule
Nesso-1        & $-0.130$ & $-5.12$ & $\mathbf{-0.135}$ & $\mathbf{-5.22}$ & $+0.023$ & $+0.62$ \\
Boltz-2        & $-0.152$ & $-3.39$ & $\mathbf{-0.185}$ & $\mathbf{-4.42}$ & $+0.132$ & $+2.08$ \\
Chai-1 (ipTM)  & $-0.148$ & $-4.12$ & $-0.138$ & $-3.80$ & $-0.043$ & $-0.84$ \\
gnina          & $-0.072$ & $-2.94$ & $-0.067$ & $-2.87$ & $-0.026$ & $-1.00$ \\
smina          & $+0.031$ & $+3.39$ & $+0.031$ & $+3.39$ & $-0.005$ & $-0.47$ \\
RF-QSAR        & $+0.002$ & $+0.05$ & $-0.009$ & $-0.31$ & $+0.067$ & $+3.50$ \\
ligand-$k$NN   & $-0.004$ & $-0.20$ & $-0.008$ & $-0.43$ & $+0.027$ & $+1.72$ \\
\bottomrule
\end{tabular}
\end{table}

\paragraph{The support curve among targets whose exact sequence never repeats.}
Table~\ref{tab:expcurve} recomputes the matched-$\mathrm{p}K$ curve on the $E=0$ subset. The singleton
bin is entirely $E=0$ by construction, so this asks precisely the objection's question: with the exact
protein appearing nowhere else in the corpus, does homologous family support still buy accuracy?
Nesso-1's gap is undiminished by the restriction, $+0.49$ against $+0.45$ unrestricted, with an
interval excluding zero, and every control stays flat under the identical restriction. Boltz-2 points
the same way but with $n=17$ in the top cell, so we report it as directional only.

\begin{table}[!ht]
\centering\footnotesize
\caption{Matched-$\mathrm{p}K$ Pearson $r$ by support bin restricted to complexes with no
identical-sequence sibling ($E=0$); cell $n$ in parentheses for the two co-folders. Two-level bootstrap
95\% CI on $G_m$.}
\label{tab:expcurve}
\setlength{\tabcolsep}{4pt}
\resizebox{\textwidth}{!}{%
\begin{tabular}{llcccccc c}
\toprule
Model & subset & 1 & 2--5 & 6--20 & 21--80 & 81--300 & 301+ & $G_m$ (95\% CI) \\
\midrule
Nesso-1 & $E{=}0$ & $+0.11$ (375) & $+0.23$ (229) & $+0.45$ (157) & $+0.39$ (102) & $+0.30$ (105) & $\mathbf{+0.60}$ (90) & $\mathbf{+0.49}$ $[+0.23,+0.74]$ \\
Nesso-1 & all     & $+0.11$ & $+0.22$ & $+0.38$ & $+0.42$ & $+0.44$ & $+0.55$ & $+0.45$ $[+0.27,+0.61]$ \\
Boltz-2 & $E{=}0$ & $-0.03$ (64) & $-0.08$ (46) & $+0.10$ (24) & $+0.78$ (16) & $+0.31$ (13) & $+0.92$ (17) & $+0.95$ $[+0.53,+1.28]$ \\
Boltz-2 & all     & $-0.03$ & $-0.01$ & $-0.01$ & $+0.42$ & $+0.37$ & $+0.59$ & $+0.62$ $[+0.13,+0.99]$ \\
smina        & $E{=}0$ & $+0.07$ & $+0.14$ & $+0.16$ & $+0.05$ & $+0.02$ & $+0.15$ & $+0.08$ $[-0.11,+0.22]$ \\
RF-QSAR      & $E{=}0$ & $+0.24$ & $+0.22$ & $+0.32$ & $+0.33$ & $+0.21$ & $+0.31$ & $+0.07$ $[-0.11,+0.22]$ \\
ligand-$k$NN & $E{=}0$ & $+0.24$ & $+0.18$ & $+0.19$ & $+0.29$ & $+0.18$ & $+0.26$ & $+0.03$ $[-0.15,+0.19]$ \\
gnina        & $E{=}0$ & $+0.08$ & $+0.09$ & $+0.19$ & $+0.14$ & $+0.29$ & $+0.18$ & $+0.09$ $[-0.26,+0.47]$ \\
\bottomrule
\end{tabular}}
\end{table}

\paragraph{Pushing the stringency further.} Table~\ref{tab:expstring} repeats the coarse novel-versus-
redundant contrast with ``no sibling'' tightened from exact match to 99\% and 95\% identity clusters.
The point estimate is stable across stringencies; only the interval widens, because at 95\% identity
just $18$ redundant-family complexes remain corpus-wide. We state that as a power limit, not as
evidence of absence.

\begin{table}[!ht]
\centering\footnotesize
\caption{Novel ($S_f\le5$) versus redundant ($S_f\ge81$) contrast under increasingly strict
definitions of ``no sibling''.}
\label{tab:expstring}
\begin{tabular}{llccc c}
\toprule
Model & no sibling at & $r$ novel & $r$ redundant & $\Delta$ & 95\% CI \\
\midrule
Nesso-1      & exact  & $+0.16$ (604) & $+0.44$ (195) & $\mathbf{+0.28}$ & $\mathbf{[+0.02,+0.48]}$ \\
Nesso-1      & 99\% id & $+0.12$ (478) & $+0.37$ (36) & $+0.25$ & $[-0.18,+0.73]$ \\
Nesso-1      & 95\% id & $+0.14$ (451) & $+0.51$ (18) & $+0.37$ & $[-0.79,+0.82]$ \\
smina        & exact  & $+0.09$ & $+0.09$ & $-0.00$ & $[-0.15,+0.09]$ \\
RF-QSAR      & exact  & $+0.23$ & $+0.27$ & $+0.04$ & $[-0.09,+0.15]$ \\
ligand-$k$NN & exact  & $+0.20$ & $+0.23$ & $+0.03$ & $[-0.09,+0.15]$ \\
\bottomrule
\end{tabular}
\end{table}

\paragraph{What this licenses, and what it does not.} The family-support gradient is not exact-target
repetition. It holds at full strength among targets whose exact sequence appears nowhere else in the
corpus, and the family-disjoint and empirical-scoring controls stay flat under the identical
restriction, so ``protein-family familiarity'' is the right framing for Nesso-1 and the paper does not
need to retreat to ``target and family redundancy''. Three things are not licensed and we state them.
First, $E$ is defined \emph{within the MIRAGE corpus}, so this rules out repetition inside the
benchmark's own corpus, which is the source of $S_f$, and not exposure to the same target from other
databases. Second, the restricted test is powered for Nesso-1 and directionally for Boltz-2 and
Chai-1, but not for gnina, whose restricted $G_m$ is $+0.09$ $[-0.26,+0.47]$; gnina's family
dependence rests on its regression slope, which holds with $E$ included ($t=-2.87$), not on a
restricted $G_m$. Third, the $301{+}$ bin comprises seven families whether or not the restriction is applied, a property
of PDBbind, which is why the family-level bootstrap is the reported interval and why the
continuous-slope regression over $2{,}203$ family clusters carries much of the argument. That said, the
endpoint is not carried by any one of those families either (Table~\ref{tab:loo}).

\section{The family-identity baseline and modality ablation}
\label{app:modality}

How much of a random-split score can family identity alone produce? A predictor that ignores its
inputs entirely and outputs the mean $\mathrm{p}K$ of the target's family scores Pearson
$\mathbf{0.564}$ under random splitting, close to Nesso-1's pooled $0.588$, and, under family-disjoint splitting, a constant prediction whose Pearson correlation is undefined; we
report it as zero by convention. This is a family-identity baseline, not a ceiling: the full ligand-plus-protein model reaches $0.740$
(Table~\ref{tab:modality}), so information beyond target identity is clearly being used. What the
baseline bounds is \emph{interpretation}, not performance: on a benchmark where target identity alone
reaches $0.56$, a reported $0.59$ cannot be read as evidence of modelled interaction physics without
decomposing it. Each test complex's family mean is computed from training-fold labels only; including
its own label would inflate the baseline to $0.670$.

\begin{figure}[!ht]
\centering
\includegraphics[width=0.62\linewidth]{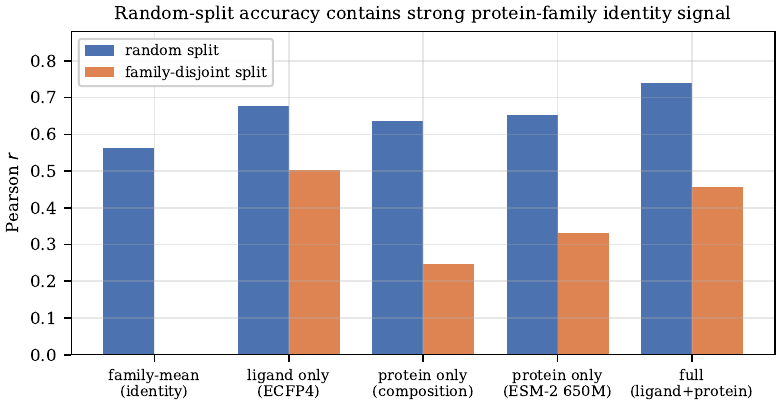}
\caption{Random-split accuracy contains strong protein-family identity signal: a protein-only model
nearly matches the full model under random splitting and collapses under family-disjoint splitting.}
\label{fig:modality}
\end{figure}

The modality ablation (Table~\ref{tab:modality}, Figure~\ref{fig:modality}) makes the same point with
a model that is not degenerate. A random forest given \emph{only the protein}, and therefore blind to the ligand
and in principle unable to know which of several ligands binds a given target more tightly, reaches
$r=0.637$ from amino-acid composition alone, or $0.653$ from ESM-2 650M embeddings
\citep{rives2021esm}, under random splitting. Under family-disjoint splitting the same models collapse
to $0.248$ and $0.330$. The ligand-only model, by contrast, loses much less ($0.676\to0.503$): protein features lose
substantially more predictive signal under family-disjoint evaluation than ligand descriptors do.

\begin{table}[!ht]
\centering\small
\caption{Modality ablation and identity baseline (random forest, Pearson $r$). A protein-only model, blind to the ligand, nearly
matches the full model under random splitting but collapses under family-disjoint splitting. The gap column quantifies the loss associated with enforcing family-disjoint evaluation, not a causal
decomposition of the score.}
\label{tab:modality}
\begin{tabular}{lccc}
\toprule
Features & Random split & Family-disjoint split & Gap \\
\midrule
family-mean (identity)       & 0.564 & $\approx 0$ & -- \\
ligand only (ECFP4)          & 0.676 & 0.503 & 0.17 \\
protein only (composition)   & 0.637 & 0.248 & \textbf{0.39} \\
protein only (ESM-2 650M)    & 0.653 & 0.330 & 0.32 \\
full (ligand $+$ protein)    & 0.740 & 0.456 & 0.28 \\
\bottomrule
\end{tabular}
\end{table}

\section{External low-support temporal evaluation}
\label{app:temporal}

\begin{figure}[!ht]
\centering
\includegraphics[width=0.62\linewidth]{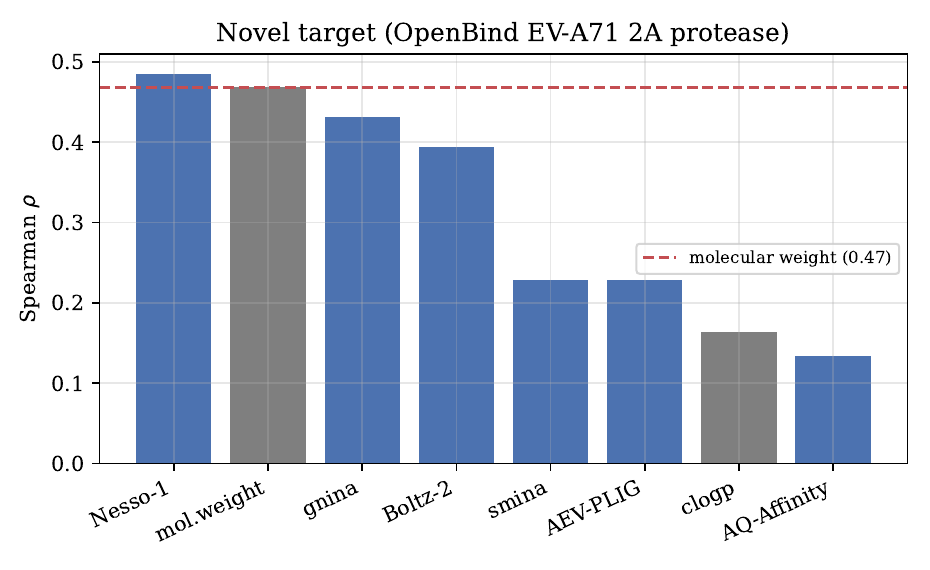}
\caption{External low-support ranking on the OpenBind EV-A71 2A protease. The dashed line is the
molecular-weight baseline; only one method clears it, and not convincingly.}
\label{fig:temporal}
\end{figure}

The external temporal arm asks whether the low-support result reproduces on a newly released ligand
series, and we state the target's provenance precisely because an earlier draft overstated it. The picornaviral 2A protease
fold is \emph{not} absent from the pre-cutoff PDB: querying RCSB for CVA16 and EV-A71 2A protease
entries returns 4MG3 (CVA16 2A protease, released 2014-03-26) and 7DA6 (EV-A71 2A protease,
2021-08-25) before 2021-09-30, no entries at all between 2021-09-30 and 2023-06-01, and $472$ entries
afterwards, all from the OpenBind/ASAP campaign. What post-dates every documented cutoff is therefore
the \emph{ligand series and its affinity measurements}, not the protein. Two public structures of this family therefore predate the documented cutoffs. We report this as
``two prior public family structures'' rather than as $S_f=2$: the count comes from the RCSB audit
above, whereas $S_f$ is defined on the PDBbind reference corpus used for the redundancy arm, which
ends in 2019 and admits only complexes carrying an affinity label, so 7DA6 could not enter it. The two
quantities sit on the same conceptual axis but not on the same reference set, and we do not equate
them. This places the target at the low-support end of the curve, so the arm is an external instance of exactly the regime the redundancy arm predicts rather than a
claim of structural novelty that the PDB contradicts. The arm therefore tests performance on a target with almost no pre-cutoff \emph{structural} support.
Because no evaluated model discloses an affinity-data cutoff (Table~\ref{tab:cutoffs}), we cannot
establish affinity-label priority for any model, and do not claim it. We describe the arm as a
zero-shot temporal evaluation rather than a prospective one: OpenBind was released publicly on 2026-05-05, and although
the Nesso-1 report describes its own OpenBind evaluation as zero-shot, we cannot establish for every
evaluated model that its weights were frozen before the measurements became available, particularly
given that no model discloses an affinity-data cutoff (Table~\ref{tab:cutoffs}).
Table~\ref{tab:temporal} and Figure~\ref{fig:temporal} report within-target Spearman $\rho$ over the
$643$ compounds with complete predictions, together with $\rho$ after partialling out molecular
weight.

Molecular weight alone reaches $\rho=0.469$. Nesso-1 exceeds it only marginally ($0.486$), Boltz-2
falls below it ($0.395$), and Gnina, smina, AEV-PLIG, clogp and AQ-Affinity all fall below it as well.
The ``beyond molecular weight'' column is the more diagnostic one: several methods are
\emph{negatively} informative once size is removed, meaning their apparent ranking skill on this
target was size tracking. Two external sources agree with this reading: Nesso-1's own report states
that its advantage over molecular weight on this target is not statistically convincing
\citep{valencelabs2026nesso1}, and the OpenBind release itself flags molecular weight as a strong
baseline for the campaign \citep{openbind2026dataset}. We report this as one external target, not as
population-level proof; a single congeneric series cannot carry that weight, and we say so in the body.

\begin{table}[!ht]
\centering\small
\caption{Novel-target ranking (OpenBind EV-A71 2A protease, $643$ compounds). Spearman $\rho$, and
$\rho$ after removing molecular weight.}
\label{tab:temporal}
\begin{tabular}{lcc}
\toprule
Method & Spearman $\rho$ & $\rho$ beyond mol.\ weight \\
\midrule
Nesso-1            & \textbf{0.486} & 0.233 \\
molecular weight   & 0.469 & -- \\
Gnina              & 0.431 & $-0.082$ \\
Boltz-2            & 0.395 & 0.112 \\
smina              & 0.229 & $-0.009$ \\
AEV-PLIG           & 0.228 & $-0.143$ \\
clogp              & 0.163 & $-0.057$ \\
AQ-Affinity        & 0.134 & $-0.019$ \\
\bottomrule
\end{tabular}
\end{table}

\section{Confidence-proxy co-folders}
\label{app:proxy}

Chai-1 and ESMFold2 predict structure and an interface confidence (ipTM) rather than affinity, and the
Chai-1 result is the one that matters here: a model with \emph{no affinity head} carries the family-support dependence, which is consistent with
the effect residing in the learned structural representation rather than only in a regression head
fitted to affinity labels.

Chai-1's ipTM (pooled $r=0.394$, $n=598$) \emph{itself} carries the family-support dependence, rising
from $0.155$ on novel families to $0.461$ on redundant ones, with a significant family-support slope ($t=-3.4$). Since Chai-1 has no affinity head, this is consistent with the dependence residing in the learned \emph{structural representation} rather
than only in a regression head fitted to affinity labels, though ipTM is a confidence score and is
itself responsive to difficulty. The dependence is therefore not confined to a trained affinity head, and changing the affinity readout
alone cannot be assumed to remove it. ESMFold2's ipTM is a weak and noisy proxy (pooled $0.183$, $n=599$) with no significant
family-support dependence ($t=-0.5$), so the effect is not universal across confidence proxies; that
arm is thin and we draw nothing further from it. AlphaFold3 would belong in this category but its
weights are request-only.

\begin{table}[!ht]
\centering\small
\caption{Confidence-proxy co-folders (interface ipTM scored as an affinity proxy). Weaker than a
trained affinity head, but the well-powered Chai-1 proxy shows the same family-support dependence.}
\label{tab:proxy}
\begin{tabular}{lccccc}
\toprule
Method & $n$ & pooled $r$ & novel-fam.\ $r$ & redundant-fam.\ $r$ & slope ($t$) \\
\midrule
Chai-1 (ipTM)   & 598 & 0.394 & 0.155 & 0.461 & $-3.4$ \\
ESMFold2 (ipTM) & 599 & 0.183 & 0.102 & 0.022 & $-0.5$ \\
\bottomrule
\end{tabular}
\end{table}

\section{Docking scoring functions: gnina depends on family support, smina does not}
\label{app:docking}

Docking methods produce an affinity-correlated score, so they enter MIRAGE directly with no change to
the protocol. Re-scoring the crystal pose separates two regimes cleanly (Table~\ref{tab:docking}).

The classical empirical score (smina/AutoDock Vina) is a weak affinity predictor overall ($r=0.095$)
and, critically, shows \emph{no} improvement with family support: its slope is significantly positive
($t=+2.4$), like molecular weight's, meaning error rises slightly as families become better
represented. We state this as measured rather than as flatness: a significant positive slope is not
flat, and the claim the evidence supports is that smina does not exhibit the performance improvement
with family support seen in the learned models. We offer no mechanism for the sign; a globally fitted
function does not need family-specific parameters in order to encode family-correlated regularities,
so the observation stands on its own. Note also that smina \citep{koes2013smina} and Vina
\citep{trott2010vina} are regression-fitted empirical functions, not free-energy calculations, which
is why we do not describe this contrast as ``physics versus learned''. This is the reference against which the co-folders' negative slopes should be read. gnina's CNN is a much stronger predictor ($r=0.444$) and carries the same significant dependence as the co-folders ($t=-2.9$), with its skill concentrated on families represented in its
PDBbind training set.

The consequence for how the field should think about this: the family-support gradient is not a
property of \emph{co-folding} versus \emph{docking}, or of large models versus small ones. The
contrast extends beyond co-folders: gnina exhibits the same family-support dependence, whereas smina
does not. Two scorers do not establish a law about every learned function, and we do not state one.

\begin{table}[!ht]
\centering\small
\caption{Docking scoring functions on MIRAGE (affinity). The classical empirical score does not
improve with family support; a CNN score trained on PDBbind shows the same family-support dependence
as the co-folders.}
\label{tab:docking}
\begin{tabular}{llcccc}
\toprule
Scorer & type & pooled $r$ & novel-fam.\ $r$ & redundant-fam.\ $r$ & slope ($t$) \\
\midrule
smina (Vina)        & classical empirical & 0.095 & 0.167 & 0.061 & $+0.018\ (+2.4)$ \\
gnina (CNNaffinity) & learned             & 0.444 & 0.330 & 0.423 & $-0.060\ (-2.9)$ \\
\bottomrule
\end{tabular}
\end{table}

\section{Pose prediction: MSA ablation, per-complex effects, and hardware}
\label{app:pose}

Table~2 of the body gives the pose results on the full $50+50$ target split. This appendix adds the
matched MSA comparison, the per-complex direction of the effect, and the practical notes needed to
reproduce the runs.

\paragraph{Pose prediction: methodology.} The same family axis applies to geometry. We redock the
crystal ligand into $50$ singleton and $50$ redundant
($S_f{\ge}301$) targets, score symmetry-corrected RMSD, count a target recovered at RMSD $<2$\,\AA{}
and a missing pose as a failure, and define pose-$G_m$ as redundant minus novel success. Five engines span the space: classical empirical docking (smina), two generative neural dockers (SigmaDock,
DiffDock-L), and two co-folders that build the whole complex de novo rather than docking into a given
receptor (Chai-1, Boltz-2), each run with and without a ColabFold MSA. All ran on Blackwell (sm\_120) hardware; DiffDock-L needed only a
PyTorch~2.7/CUDA~12.8 environment, not code changes.

\begin{table}[!ht]
\centering\footnotesize
\caption{Pose prediction (redocking; success $=$ RMSD $<2$\,\AA, $n{=}50$ per bin, no pose $=$
failure). The two blocks receive different information and are not comparable across the divide.
Chai-1's MSA comparison, which requires a matched subset, is in Appendix~\ref{app:pose}.}
\label{tab:pose}
\setlength{\tabcolsep}{6pt}
\begin{tabular}{llccc}
\toprule
Engine & input & novel success & redundant success & pose-$G_m$ \\
\midrule
\multicolumn{5}{l}{\emph{given the crystal receptor}: smina sets a $+0.23$ difficulty reference for this block} \\
smina      & classical empirical docking & 0.62 & 0.86 & $+0.23$ \\
SigmaDock  & neural diffusion docking    & 0.52 & 0.80 & $+0.28$ \\
DiffDock-L & neural diffusion docking    & 0.18 & 0.54 & $+0.36$ \\
\midrule
\multicolumn{5}{l}{\emph{building the complex de novo}} \\
Chai-1     & co-folding, no MSA & 0.28 & 0.64 & $+0.36$ \\
Boltz-2    & co-folding, no MSA & 0.02 & 0.66 & \textbf{+0.64} \\
Boltz-2    & co-folding, $+$MSA & 0.46 & 0.66 & \textbf{+0.20} \\
\bottomrule
\end{tabular}
\end{table}

\paragraph{Matched MSA comparison.} Table~\ref{tab:posemsa} compares each co-folder with and without a
ColabFold MSA. Boltz-2 is the full $n=50$ per bin. Chai-1 did not complete every setting, so its rows
use the matched subset both settings finished ($36$ novel, $33$ redundant), which is why its no-MSA
numbers differ slightly from the body table's full-set values. The two co-folders do \emph{not} behave the same way. For Boltz-2 the MSA lifts novel-family success
and leaves redundant families exactly unchanged, collapsing the gap from the novel side. For Chai-1,
novel success rises from $22\%$ to $33\%$ but redundant success falls from $58\%$ to $46\%$: of the
$0.23$ reduction in pose-$G_m$, $+0.11$ is novel-family gain and $-0.12$ is redundant-family loss, so
over half the effect is degradation rather than rescue. Chai-1's $+$MSA runs also failed to produce a
pose for $15$ of $50$ novel and $17$ of $50$ redundant targets (\texttt{n\_posed} $35$ and $33$ in
\texttt{results/pose\_chai\_msa.json}), which is why the comparison needs a matched subset at all;
Boltz-2 produced poses for all $50$ in both settings. Chai-1's redundant decline is therefore measured
on the complexes that happened to complete and cannot be cleanly separated from run attrition.

\paragraph{The effect is one-directional per complex.} Aggregate success rates could in principle hide
churn, with an MSA fixing some complexes while breaking others. It does not. Across the $50$ novel
Boltz-2 targets the MSA flips $22$ complexes from failure to success and \emph{zero} from success to
failure. Across the $50$ redundant targets it gains $3$ and loses $3$, a net of zero. Chai-1 shows the same direction on the novel families of its matched subset ($5$ gained, $1$ lost),
but its redundant bin moves the other way, so the one-directional reading holds for Boltz-2 only. For
Boltz-2 the MSA is not trading one kind of error for another; it is supplying something that was
missing, and it was only missing on unfamiliar families.

\paragraph{Interpretation.} Supplying evolutionary context removes much of the same pose deficit that is associated with low
public family support; we do not claim that family redundancy itself acts by supplying evolutionary
context during training, since training-set redundancy was never experimentally varied. This is
why we describe the co-folder pose gap as a \emph{localisable} deficit rather than irreducible
difficulty: MIRAGE identifies not only that single-sequence co-folders lean on family familiarity,
but \emph{what} they are missing when they do not have it. It is the most direct evidence in the paper that the novel-family deficit is a \emph{missing-input}
problem rather than an irreducible one: with architecture, weights and target set held fixed,
supplying evolutionary context recovers $22$ of the $49$ novel-family failures. We note what this does
not establish. No arm of this study varies training-set redundancy, so the comparison against ``more
redundant training data'' is an inference from the shape of the support curve, not a measured
contrast. And family novelty here is defined on PDBbind structural support, whereas ColabFold
retrieves from far larger sequence databases; a target can be structurally novel and still have a deep
sequence family, which is precisely why the rescue is possible. Quantifying that relationship, by
regressing MSA depth and $N_{\mathrm{eff}}$ on $\log_{10}S_f$ across the pose targets, is the natural
next measurement and the harness ships the script for it.

\begin{table}[!ht]
\centering\small
\caption{Effect of a ColabFold MSA on co-folder pose recovery. Boltz-2 shows a pure novel-family
rescue: novel success rises while redundant success is unchanged, on the full $n{=}50$ per bin. Chai-1
shows a mixed effect on the matched subset both settings completed ($36$ novel, $33$ redundant),
gaining on novel families and losing on redundant ones.}
\label{tab:posemsa}
\begin{tabular}{llcccc}
\toprule
Co-folder & input & novel success & redundant success & pose-$G_m$ & flips (fail$\to$OK / OK$\to$fail) \\
\midrule
Boltz-2 & no MSA & 0.02 & 0.66 & $+0.64$ & -- \\
Boltz-2 & $+$MSA & 0.46 & 0.66 & \textbf{+0.20} & $22 / 0$ \\
Chai-1  & no MSA & 0.22 & 0.58 & $+0.35$ & -- \\
Chai-1  & $+$MSA & 0.33 & 0.46 & \textbf{+0.12} & $5 / 1$ \\
\bottomrule
\end{tabular}
\end{table}

\paragraph{Hardware and environment notes.} All pose models were run on current-generation Blackwell
(sm\_120) hardware. This required adapting DiffDock-L, whose released environment pins CUDA-11.7-era
PyTorch and therefore ships no sm\_120 kernels: installing PyTorch 2.7 with CUDA 12.8 and matching
pre-built PyTorch-Geometric wheels, and dropping the unused OpenFold dependency, makes it run unmodified. No custom CUDA kernels are involved, so this is a packaging fix rather
than a port. Two
methods remained impractical on this hardware and are excluded rather than reported weakly: Uni-Mol
Docking V2 \citep{yu2023unidock}, whose Uni-Core dependency is frozen at CUDA 11.8, and gnina's
CNN-refinement docking mode, which is prohibitively slow at this scale (its rescoring mode is used for
affinity in Appendix~\ref{app:docking}). MIRAGE ships the pose harness together with the DiffDock,
SigmaDock and Chai-1/Boltz-2 adapters, the last two in both MSA settings, so further engines plug in
directly.

\section{Family-balanced sensitivity: does the estimand change the conclusion?}
\label{app:balanced}

The estimator of Appendix~\ref{app:defs} weights every complex equally, so a family contributing
hundreds of ligands influences a bin's $r$ far more than a family contributing one. In the $301{+}$
bin, where thousands of complexes come from seven families, this is not a hypothetical concern: the
reported $r$ estimates performance for a random \emph{complex} drawn from that bin, whereas the paper
interprets MIRAGE as a statement about new \emph{targets and families}. The two-level bootstrap makes
the interval family-aware but leaves the estimand complex-weighted, so we recompute both quantities
under a family-balanced estimator in which each complex carries weight $w_i=1/n_{f(i)}$ within its stratum, giving every family equal total mass.
Here $n_f$ is the number of \emph{analysed} complexes from family $f$, that is those remaining after
the method's prediction coverage and the matched $\mathrm{p}K$ restriction, counted within the
stratum. This is materially smaller than the corpus family size: the $S_f{\ge}301$ stratum holds
$3{,}970$ corpus complexes but only $336$ in Nesso-1's matched analysis cell, and it is over those
$336$ that the seven families are equalised. This is well defined in the singleton bin, where every
weight is one, which is why we prefer it to per-family correlations (undefined there) and to a
leave-one-high-support-family-out check (which addresses influence rather than the estimand).

\paragraph{The gap under both estimands.} Table~\ref{tab:balanced} reports $G_m$ complex-weighted and
family-balanced. Nesso-1 is essentially unchanged, $+0.45$ against $+0.44$ with an interval excluding
zero, and every control stays at or near zero under both. The shift is small and lands exactly where
the objection predicts it should: Nesso-1's $301{+}$ bin holds $336$ complexes from seven families,
the largest supplying a quarter of them, and equalising those seven moves the bin from $0.551$ to
$0.541$. The singleton bin is identical by construction, so that $0.010$ is the whole of the
difference in $G_m$.

\begin{table}[!ht]
\centering\footnotesize
\caption{$G_m$ under the complex-weighted and family-balanced estimators (matched $\mathrm{p}K$
window; $w_i=1/n_{f(i)}$ within stratum). Nesso-1's gap is insensitive to the estimand; Boltz-2's
interval spans zero at its current coverage.}
\label{tab:balanced}
\begin{tabular}{lccc}
\toprule
Method & complex-weighted & family-balanced & 95\% CI (balanced) \\
\midrule
Nesso-1      & $+0.45$ & $\mathbf{+0.44}$ & $[+0.26,+0.61]$ \\
Boltz-2      & $+0.62$ & $+0.37$ & $[-0.24,+1.00]$ \\
gnina        & $+0.07$ & $+0.03$ & $[-0.20,+0.29]$ \\
smina        & $-0.04$ & $-0.03$ & $[-0.17,+0.22]$ \\
RF-QSAR      & $-0.01$ & $-0.04$ & $[-0.18,+0.15]$ \\
ligand-$k$NN & $-0.03$ & $-0.03$ & $[-0.17,+0.11]$ \\
\bottomrule
\end{tabular}
\end{table}

\paragraph{The continuous slope under inverse-family-size weights.} Table~\ref{tab:balancedreg}
repeats the covariate regression as weighted least squares with the same $1/n_{f}$ weights, CR1
clustered on family. Nesso-1's coefficient grows in magnitude when heavily sampled families are
weighted down, $-0.130$ to $-0.148$, and gnina's does the same. The controls stay flat. Down-weighting
the medicinal-chemistry families strengthens the effect rather than dissolving it, which is the
opposite of what a few-families explanation predicts.

\begin{table}[!ht]
\centering\footnotesize
\caption{Covariate regression of $|\mathrm{error}|$ on $\log_{10}S_f$, unweighted and with
$w_i=1/n_{f(i)}$ (WLS). CR1 standard errors clustered on family throughout.}
\label{tab:balancedreg}
\begin{tabular}{lcccc}
\toprule
Method & coef (unweighted) & $t$ & coef ($1/n_f$) & $t$ \\
\midrule
Nesso-1      & $-0.130$ & $-5.12$ & $\mathbf{-0.148}$ & $\mathbf{-4.67}$ \\
Boltz-2      & $-0.152$ & $-3.39$ & $-0.130$ & $-2.45$ \\
gnina        & $-0.072$ & $-2.94$ & $-0.088$ & $-2.57$ \\
smina        & $+0.031$ & $+3.39$ & $+0.030$ & $+2.66$ \\
RF-QSAR      & $+0.002$ & $+0.05$ & $+0.003$ & $+0.16$ \\
ligand-$k$NN & $-0.004$ & $-0.20$ & $+0.020$ & $+1.12$ \\
\bottomrule
\end{tabular}
\end{table}

\paragraph{Implementation note.} Weights are fixed from the observed stratum rather than recomputed
inside each bootstrap replicate. Recomputing them would give a family drawn twice the same total mass
as a family drawn once, which would erase precisely the family-level variability the bootstrap exists
to measure.

\section{LBA30: the missing control on a standard benchmark}
\label{app:lba}

ATOM3D LBA \citep{townshend2021atom3d} is the field's standard structure-based affinity-generalization
benchmark: PDBbind-refined complexes split at 30\% protein sequence identity (``LBA30''; 3507 train /
490 test), on which recent geometric deep networks report large gains. IPBind \citep{li2025ipbind}
reports Pearson $0.732$, a claimed 19.6\% relative improvement. No published LBA30 result we could
find reports a ligand-only, molecular-weight or QSAR control; the comparison is always against other
structure or sequence deep networks. We supply that control in Table~\ref{tab:lba30}. The official split is 3507 train / 490 test, but only
2961 and 453 entries respectively carry a usable SMILES field. We recover the rest from the ligand
atom block: molecular weight is summed from the element list and is therefore defined on all 490,
matching the $n$ of the published deep-net numbers, and bond perception supplies ECFP4 and clogp for
all but one test entry. Reporting a ligand-only control on a silently reduced subset would not be a
like-for-like comparison with the leaderboard, which is the failure mode this section is about.

Two conclusions follow, and we state both. First, the top methods genuinely learn structural signal:
IPBind exceeds our ligand-only RF by roughly $0.29$, which is strong evidence for substantial
predictive signal beyond this ligand-only baseline. Second, and more consequential for the field, molecular weight \emph{alone} scores
$0.436$, above ENN ($0.389$), near DeepDTA ($0.472$), and within $0.11$ of 3DCNN ($0.550$) and GNN
($0.545$). Small reported margins over those methods were never contextualised against a trivial
descriptor, and a reader of the published leaderboard has no way to tell which entries are above the
trivial baseline and which are not.

The same control also reconciles LBA30 with MIRAGE. A 30\%-\emph{sequence} split is a far weaker
generalization test than singleton-family stratification: a ligand-only model reaches $0.440$ on LBA30 but only $0.31$--$0.41$ on MIRAGE novel families. So $0.73$ (LBA30) and $0.31$ (MIRAGE novel families)
are not in contradiction: they measure different stringencies, and only the second resembles the
situation of a genuinely new target.

A related benchmark, CORDIAL \citep{brown2025cordial}, is explicitly designed to exclude fold and
family identity and is validated leave-superfamily-out; it too reports no ligand-only or trivial
baseline. The absence of such controls across the generalization-benchmark literature is precisely the
gap MIRAGE institutionalises: not a criticism of any one paper, but of a reporting convention that
lets a trivial descriptor go unmeasured.

\begin{table}[!ht]
\centering\small
\caption{LBA30, full official test split ($n{=}490$; $489$ for the fingerprint-based controls),
Pearson $r$ on $\mathrm{p}K$. Structure/sequence deep-net numbers are from the literature
\citep{townshend2021atom3d,li2025ipbind}; the three ligand-only controls (bold) are ours and, to our
knowledge, have not previously been reported.}
\label{tab:lba30}
\begin{tabular}{lc@{\hskip 3em}lc}
\toprule
Structure/sequence models & $r$ & Ligand-only controls (ours) & $r$ \\
\midrule
IPBind      & 0.732 & \textbf{ligand-only RF (ECFP4)} & \textbf{0.440} \\
EHIGN       & 0.612 & \textbf{molecular weight}       & \textbf{0.436} \\
GIGN        & 0.586 & \textbf{ligand-$k$NN}           & \textbf{0.417} \\
3DCNN       & 0.550 & clogp                            & 0.202 \\
GNN         & 0.545 & & \\
DeepDTA     & 0.472 & & \\
ENN         & 0.389 & & \\
\bottomrule
\end{tabular}
\end{table}

\section{Extended discussion and limitations}
\label{app:discussion}

\paragraph{Why we avoid the word ``leakage''.} Our evidence establishes \emph{public family-support dependence} and \emph{redundancy-driven
inflation}: accuracy scales with public family support after controlling for measured confounds, and matched family-disjoint models do not share the
dependence. We deliberately avoid ``leakage'', which should be reserved for evaluations where related
families demonstrably occur across a claimed train/test boundary. We cannot reconstruct the exact
proprietary affinity training sets of the frontier co-folders, and rather than speculate we make the
weaker, defensible claim: public affinity benchmarks are dominated by families with high public structural support, on which the
evaluated models achieve substantially higher accuracy, and family identity alone reproduces a large
fraction of the headline correlation, so the pooled correlation
cannot be interpreted as evidence of transferable interaction modelling.
The distinction is not pedantry. A leakage claim would be a claim about someone's data pipeline and
would be answerable by fixing the split; the claim we make is about the benchmark distribution itself
and is not addressed by re-splitting alone. A family-disjoint split with retraining estimates transfer
directly; MIRAGE audits frozen models across the support axis, which is complementary.

\paragraph{What follows for practice.} Co-folders are the most accurate methods available on
well-supported families, and nothing here contradicts that. The problem is the standard practice of
reporting one overall correlation, which conflates interpolation within familiar families with
transfer to novel ones. For the evaluated models, the low-support values of roughly $0.1$--$0.3$ are more relevant to a new
target than their pooled correlations, and a cheap family-disjoint QSAR model is a strong, rarely-reported baseline that costs seconds to run
and, in that regime, wins.

\paragraph{Within-target ranking is not estimable on this corpus.} A natural alternative to
cross-target correlation is within-exact-target Spearman, which is what a medicinal chemistry campaign
optimises. MIRAGE cannot supply it: requiring at least five ligands against an identical protein
sequence leaves six targets ($69$ ligands) for Nesso-1 in the redundant stratum and \emph{none} in the
novel stratum, and Boltz-2 has none in either (Table~\ref{tab:withintarget}). This is a limitation of
PDBbind-derived coverage rather than of the metric, and it is the strongest argument for a future arm
built from several ligands per low-support target with support defined on a separate structural
corpus.

\begin{table}[!ht]
\centering\footnotesize
\caption{Within-exact-target Spearman, requiring $\ge5$ ligands per target. The novel stratum supports
no such estimate for any model.}
\label{tab:withintarget}
\begin{tabular}{llccc}
\toprule
Model & stratum & targets & ligands & mean $\rho$ \\
\midrule
Nesso-1      & redundant & 6  & 69   & $+0.593$ \\
gnina        & redundant & 6  & 69   & $+0.442$ \\
RF-QSAR      & redundant & 64 & 1023 & $+0.141$ \\
ligand-$k$NN & redundant & 64 & 1023 & $+0.106$ \\
smina        & redundant & 64 & 1022 & $+0.139$ \\
any model    & novel ($S_f\le5$) & 0--1 & $\le5$ & not estimable \\
\bottomrule
\end{tabular}
\end{table}

\paragraph{Limitations.} The redundancy set is pre-2020 PDBbind, and $S_f$ is a within-corpus support proxy rather than the
models' true training counts; high-confidence structural cutoffs are documented for only two evaluated models, with a
medium-confidence time split available for DiffDock-L (Table~\ref{tab:cutoffs}), and no model
discloses a cutoff for its affinity training data; a model trained on a corpus with different
family structure would need its own support annotation. The temporal set is a single target with a
congeneric series, which is a weak base for population-level claims and is reported as such. We
evaluate frozen public weights and do not retrain on de-leaked data, so we measure the models as a
practitioner would obtain them, not the best achievable version of each architecture. Support annotation at 20--50\% sequence identity changes $G_m$ by at most $0.04$ for Nesso-1 and
leaves every control flat (Table~\ref{tab:clust}); Pfam- and pocket-based clustering remain untested. Exact-target exposure is measured within the MIRAGE corpus; we cannot observe repetition in
proprietary training sets. The confidence-proxy results use ipTM rather than a trained affinity head
and should be read as evidence about representations, not as affinity benchmarks. The Boltz-2 arm covers $583$ complexes against Nesso-1's $3{,}289$, and that thinness, not a change of
direction, is what fails three robustness checks: assay-restricted strata (Appendix~\ref{app:data}),
40--50\% clustering (Appendix~\ref{app:dose}) and family-balancing (Appendix~\ref{app:balanced}). In
each the point estimate keeps its sign while the interval opens. We therefore mark Boltz-2's gap as
directional pending wider coverage, and rest the affinity claim on the Nesso-1 arm, which supports
every one of those restrictions. Finally, the pose arm uses $50+50$ targets, which is adequate to separate a $+0.23$ reference from a $+0.64$ gap but not to rank engines whose
gaps differ by a few points.

\section{Reproducibility and released artifact}
\label{app:repro}

MIRAGE is an installable package (\texttt{pip install mirage}) with an accompanying Hugging Face
dataset. Scoring a new model requires wrapping it as \texttt{predict(sequence, smiles)}
$\to$ \texttt{float} and calling \texttt{run\_redundancy} or \texttt{run\_temporal}; a CLI scores a
predictions CSV directly, so a model that has already been run offline needs no integration at all:

\begin{quote}\small\ttfamily
mirage score redundancy predictions/nesso1\_redundancy.csv \\
mirage score redundancy predictions/rf\_qsar\_redundancy.csv \\
python scripts/train\_baselines.py -{}-out-dir predictions/
\end{quote}

All metrics and controls ($G_m$, matched-$\mathrm{p}K$ strata, the multivariate regression, the
two-level bootstrap), the full baseline suite, the pose harness with its five engine adapters, and the
dataset construction pipeline are included. Every number in this paper is reproduced by the shipped
harness from the released per-method prediction CSVs, which are versioned alongside the code so that a
future model can be compared against exactly these predictions rather than a re-run of them.
Code: \url{https://github.com/DeepBio-Scientific/MIRAGE}, pinned at the release commit and archived
with a DOI; dataset: \url{https://huggingface.co/datasets/DeepBioScientific/MIRAGE}, versioned to match.
The leave-one-family-out, Spearman, family-balanced and cross-fitted-calibration analyses are written
to \texttt{results/sensitivity\_stats.json} and \texttt{results/estimand\_stats.json} by the shipped
scripts.

\paragraph{Declarations.} Funding and competing interests are declared in the submission metadata; the
authors are affiliated with DeepBio Scientific, which does not develop any of the evaluated models.

\end{document}